\documentclass{article}

\usepackage[english]{babel}
\usepackage[letterpaper,top=2cm,bottom=2cm,left=3cm,right=3cm,marginparwidth=1.75cm]{geometry}

\usepackage{amsmath}
\usepackage{amssymb}
\usepackage{mathtools}
\usepackage{graphicx}
\usepackage[colorlinks=true, allcolors=blue]{hyperref}

\usepackage{authblk}
\usepackage{comment,xcolor}
\usepackage{paralist, tabularx}

\title{Coupled Environmental and Demographic Fluctuations Shape the Evolution of Cooperative Antimicrobial Resistance}
\author{Llu\'is Hern\'andez-Navarro*, Matthew Asker, Alastair M. Rucklidge, and Mauro Mobilia\(^\dagger\)}
\affil{Department of Applied Mathematics, School of Mathematics,\\ University of
Leeds, Leeds LS2 9JT, U.K.}
\affil{*L.Hernandez-Navarro@leeds.ac.uk, \(^\dagger\)M.Mobilia@leeds.ac.uk}

\begin{document}
\maketitle

\begin{abstract}
There is a pressing need to better understand how microbial populations respond to antimicrobial drugs, and to find mechanisms to possibly eradicate antimicrobial-resistant cells. The inactivation of antimicrobials by resistant microbes can often be viewed as a cooperative behavior leading to the coexistence of resistant and sensitive cells in large populations and static environments. This picture is however greatly altered by the fluctuations arising in volatile environments, in which microbial communities commonly evolve. Here, we study the eco-evolutionary dynamics of a population consisting of an antimicrobial resistant strain and microbes sensitive to antimicrobial drugs in a time-fluctuating environment, modeled by a carrying capacity randomly switching between states of abundance and scarcity. We assume that antimicrobial resistance is a shared public good when the number of resistant cells exceeds a certain threshold. Eco-evolutionary dynamics is thus characterised by demographic noise (birth and death events) coupled to environmental fluctuations which can cause population bottlenecks. By combining analytical and computational means, we determine the environmental conditions for the long-lived coexistence and fixation of both strains, and characterise a \emph{fluctuation-driven} antimicrobial resistance eradication mechanism, where resistant microbes experience bottlenecks leading to extinction. We also discuss the possible applications of our findings to laboratory-controlled experiments.
\end{abstract}

\section{Introduction}
Environmental conditions, such as  temperature, pH, or available resources, endlessly change over time and shape the fate of natural populations. For instance, microorganisms often live in volatile environments where resource abundance fluctuates  between mild and harsh conditions, and  regimes of feast alternate with periods of famine~\cite{Srinivasan98,Merritt18,Himeoka19}. How environmental variability (EV), generally referring to changes  not caused by the organisms themselves (e.g., supply of abiotic resources), affects species diversity is a subject of intense debate and research, see, e.g.,~\cite{Grime73,Miller11,Fox13,Chesson81,Chesson94,Chesson00,Ellner19,Murugan21,visco2010switching,gaal2010exact}. Demographic noise (DN) arising from randomness in birth and death events in finite populations is another source of fluctuations. DN is negligible in large populations and strong in small ones, where it can lead to species fixation, when one species takes over the population, or to extinction, and hence can permanently set the make-up of a community~\cite{Ewens,Kimura,Blythe07,Plotkin23}. The dynamics of the population composition (evolutionary dynamics) is often coupled with that of its size (ecological dynamics)~\cite{Roughgarden79}, resulting in its \emph{eco-evolutionary dynamics}~\cite{Pelletier09,Harrington14,wienand2018eco}.

When EV influences the size of a population, it also modulates the DN strength, leading to a coupling of DN and EV~\cite{wienand2017evolution,wienand2018eco,west2020,taitelbaum2020population,Shibasaki2021,taitelbaum2023evolutionary}. This interdependence is potentially of great relevance to understand eco-evolutionary dynamics of microbial communities. The coupling of DN and EV can lead to population bottlenecks, where new colonies consisting of few individuals are prone to fluctuations~\cite{Wahl02,Patwas09,Brockhurst07,Brockhurst07b}, and plays an important role in the eco-evolutionary dynamics of antimicrobial resistance (AMR)~\cite{Coates18,marrec2020resist}. 

The rise of AMR is a global threat  responsible for millions of deaths~\cite{oneill2016tackling}. Understanding how AMR evolves and what mechanisms can possibly eradicate the resistance to antimicrobials are therefore questions of great societal relevance and major scientific challenges. A common mechanism of antimicrobial resistance involves the production by resistant cells, at a metabolic cost, of an extra or intracellular enzyme inactivating antimicrobial drugs~\cite{davies1994inactivation,Wright05,Yurtsev13}. When the number of resistant cells exceeds a certain threshold, there are enough drug-inactivating enzymes, and the protection against antimicrobial drugs is shared with sensitive cells that can thus also resist antimicrobial drugs at no metabolic cost. However, below the resistant population threshold, only resistant microbes are protected against the drug (enzyme availability is limited and it can only inactivate the drug in the vicinity of resistant cells). AMR can hence be viewed as a thresholded cooperative behavior where widespread antimicrobial inactivation is a form of public good. This results in the spread of resistant microbes below the threshold, while sensitive cells thrive under high enzymatic concentration (above threshold). Hence, in static environments and large populations, both sensitive and resistant strains survive antimicrobial treatment and coexist in the long run~\cite{Yurtsev13, vega2014collective, meredith2015collective, bottery2016selective}. In this work, we show that this picture can be greatly altered by the joint effect of demographic and environmental fluctuations, often overlooked, but ubiquitous in microbial communities that commonly evolve in volatile environments, where they can be subject to extreme and sudden changes~\cite{Wahl02,Brockhurst07,Brockhurst07b,Patwas09,shade2012,stegen2012,Coates18}.
 
Motivated by the problem of the evolution of AMR, here we study the coupled influence of EV and DN on the eco-evolutionary dynamics of a population of two species, one antimicrobial resistant strain and the other sensitive to antimicrobials. In our model, we assume that AMR is a cooperative behavior above a certain threshold for the number of resistant microbes, and the microbial community is subject to environmental fluctuations that can cause population bottlenecks. Here, EV involves random switches of the carrying capacity, causing the population size to fluctuate, while the antimicrobial input is kept constant. We thus study how the joint effect of EV and DN affects the fixation and coexistence properties of both strains, determining under which environmental conditions either of them prevail or if they both coexist for extended periods. This allows us to identify and fully characterise a \emph{fluctuation-driven} antimicrobial resistance eradication mechanism, where environmental fluctuations generate transients that greatly reduce the resistant population and DN can then lead to the extinction of AMR. 

In the next section, we introduce the model and discuss our methods. We present our results in section~\ref{Sec:Results}, where we first describe the main properties of the ({\it in silico}) model evolving under a fluctuating environment, and then study its properties analytically. In sections \ref{Sec:PhaseDiagram} to \ref{Sec:Moran}, we analyse the population dynamics in the large population limit, and then the model's fixation properties in static environments. In section \ref{Sec:TransientDips}, we characterise the fixation and coexistence of the strains in fluctuating environments, and discuss in detail the fluctuation-driven eradication of antimicrobial resistance arising in the regime of intermediate switching. Section \ref{Sec:Discussion} is dedicated to the discussion of the influence of environmental variability on the strains fraction and abundance (section~\ref{Sec:Discussion1}), and to a review of our modeling assumptions (section~\ref{Sec:Discussion2}). Our conclusions are presented in section~\ref{Sec:Conclusions}. Technical and computational details are given at the end, in the annex supplemental material (SM).

\section{Methods \& Models}

Microbial communities generally evolve in volatile environments: they are subject to suddenly changing conditions~\cite{shade2012,stegen2012}, and fluctuations can play an important role in their evolution~\cite{Grime73,Miller11,Fox13,Chesson94,Chesson00,Coates18,Ellner19,Murugan21}. For instance, fluctuating nutrients may be responsible for population bottlenecks leading to feedback loops and cooperative behavior~\cite{Wahl02,Brockhurst07,Brockhurst07b,Patwas09}, while sensitivity to antimicrobials depends on cell density and its fluctuations~\cite{brook2004beta,Yurtsev13,vega2014collective,meredith2015collective,bottery2016selective}. Here, we study the eco-evolutionary dynamics of cooperative AMR by investigating how a well-mixed microbial community evolves under the continued application of a drug that hinders microbial growth when the community is subject to fluctuating environments. The evolutionary dynamics of the microbial community is modeled as a multivariate birth-and-death process~\cite{Gardiner,VanKampen,Blythe07}, whereas to model the fluctuating environment we assume that the population is subject to a time-varying binary carrying capacity~\cite{Bena2006,HL06,taitelbaum2020population,taitelbaum2023evolutionary,hufton2016intrinsic}.

\subsection{Microbial model}\label{Sec:Model}
We consider well-mixed co-cultures composed of an antimicrobial resistant cooperative strain (denoted by $R$) and a defector type sensitive to antimicrobials (labeled $S$), under a constant input of antimicrobial drug, inspired by a chemostat laboratory set-up. The population, of total size $N$, hence consists of \(N_R\) resistant and \(N_S\) sensitive microbes, with \(N=N_R+N_S\). Note that, since we later introduce EV as switches in the carrying capacity, the total population will fluctuate accordingly. A frequent mechanism of antimicrobial resistance relies on the production of an \emph{enzyme hydrolysing the antimicrobial drug} in their surroundings~\cite{davies1994inactivation,Wright05,Yurtsev13}. Here we assume that each \(R\) cell produces the enzyme at a constant rate, regardless of the antimicrobial concentration, which is inspired by typical lab experiments, e.g., with resistance gene-bearing plasmids~\cite{Yurtsev13, bottery2016selective}. When the number of \(R\) is high enough, the overall concentration of resistance enzyme in the medium suffices to inactivate the drug for the entire community: the enzyme hydrolyses the drug and sets it below the \emph{Minimum Inhibitory Concentration} (MIC), therefore acting as a public good and protecting \(S\) as well. This mechanism can hence lead to antimicrobial resistance as a cooperative behavior~\cite{davies1994inactivation,Wright05,Yurtsev13}, for instance, by means of the \(\beta\)-lactamase resistance enzyme for the general \(\beta\)-lactam family of antibiotics~\cite{brook2009role} (see section~\ref{Sec:Discussion2} for non-shared resistance mechanisms).
 
Here, we model this AMR mechanism by assuming that $R$ acts as a cooperative strain when the number of $R$ cells (proxy for resistance enzyme concentration) exceeds a fixed threshold \(N_{th}\), i.e., $R$ cells are cooperators when \(N_R\geq N_{th}\), while they retain for themselves the benefit of producing the protecting enzyme when \(N_R< N_{th}\)~\cite{brook2004beta,Yurtsev13,vega2014collective,meredith2015collective,bottery2016selective}. The effective regulation of public good production by means of a population threshold has been found in a number of microbial systems, see, e.g.,~\cite{Brown01,Camilli06,Brockhurst07,sanchez2013feedback,vega2014collective}, and is consistent with a slower microbial growth cycle with respect to the fast time scale of enzyme production and dispersion. In this work, we study the AMR evolution as a form of cooperative behavior under demographic and environmental fluctuations. Assuming fixed-volume fluctuating environments, the threshold for AMR cooperation is here set in terms of $R$ abundance (rather than its concentration), see section~\ref{Sec:Discussion}.

In our model, $R$ microbes have a constant birth rate independent of the biostatic drug hindering microbial growth~\cite{hughes2012selection,andersson2007biological,sanmillan2017fitness}~\footnote{Here, for simplicity, we focus on biostatic drugs that reduce growth rate of sensitive cells $S$. Biocidal drugs would increase the death rate of $S$. In fact, our choice is not particularly limiting since the effect of a same drug can be either biostatic or biocidal, depending on the concentration of cells and antimicrobial~\cite{hughes2012selection, andersson2007biological, sanmillan2017fitness}.}, with fitness \(f_R=1-s\), where \(0<s<1\) captures the extra metabolic cost of constantly generating the resistance enzyme. The birth rate of $S$ depends on the public good abundance: when \(N_R<N_{th}\), the enzyme concentration is low (below cooperation threshold) and the antimicrobial drug is above the MIC, the $S$ fitness $f_S$ is thus lower than $f_R$, with \(f_S=1-a\), where \(1>a>s\) and $a$ encodes growth rate reduction caused by the drug. When \(N_R\geq N_{th}\), the $R$ abundance is above the cooperation threshold. This triggers the AMR cooperative mechanism: the drug is inactivated (below MIC), and the $S$ birth rate, with \(f_S=1\), is then higher than that of $R$, see figure~\ref{fig:SketchAndBehavior}a. Denoting by $x\equiv N_R/N$ the fraction of $R$ in the population, here $S$ fitness is \[f_S=1-a~\theta\left[N_{th}-N_R\right]=1-a~\theta\left[x_{th}\left(N\right)-x\right],\] where \(\theta[z]\) is the Heaviside step function, defined as \(\theta[z]=1\) if (\(z>0\)) and  \(\theta[z]=0\) otherwise, and \(x_{th}\left(N\right)\equiv N_{th}/N\) is the fraction of $R$ at the cooperation threshold. The average population fitness is \(\bar{f}=f_RN_R/N+f_SN_S/N\). In this setting, this population evolves according to the multivariate birth-death process~\cite{Gardiner,VanKampen,Ewens} defined by the reactions
\begin{align}
\label{reactions}
N_{R/S}&\xrightarrow[]{T^+_{R/S}}N_{R/S}+1 \nonumber\\N_{R/S}&\xrightarrow[]{T^-_{R/S}}N_{R/S}-1,\end{align}
occurring with transition rates~\cite{wienand2017evolution,wienand2018eco,taitelbaum2020population,Shibasaki2021}
\begin{align}
 T^+_{R}&=\frac{f_{R}}{\bar{f}}N_{R}=\frac{(1-s)~N_R}{1-a\theta\left[N_{th}-N_R\right]+(a\theta\left[N_{th}-N_R\right]-s)N_R/N}, \qquad  
T^-_{R}=\frac{N}{K}N_{R} \qquad \text{and}\qquad \nonumber\\
 T^+_{S}&=\frac{f_{S}}{\bar{f}}N_{S}=\frac{(1-a\theta\left[N_{th}-N_R\right])~N_S}{1-a\theta\left[N_{th}-N_R\right]+(a\theta\left[N_{th}-N_R\right]-s)N_R/N}, \qquad
T^-_{S}=\frac{N}{K}N_{S},
\label{eq:transrates}
\end{align}
with growth limited by the logistic death rate $N/K$ (so that the total population \(N\) follows the standard logistic dynamics in the mean field limit, see equation \eqref{det_eq_N}), where $K$ is the carrying capacity, that is here assumed to be a time-fluctuating quantity, see below. Moreover, we have normalised $f_{R/S}$ by the average fitness \(\bar{f}\) for mathematical convenience,  without loss of generality (see section~\ref{Sec:Discussion2}). This corresponds to the growth rate of each strain to be given by its fitness relative to the average population's fitness, a common assumption in the context of  biological and evolutionary processes~\cite{Ewens,Chesson81,traulsen2009stochastic}, which allows us to establish a neat relationship between our multivariate birth-death process and the classical Moran process. The latter is the reference birth-death-like process used to model the evolution of idealised populations of constant total size~\cite{Moran, Ewens, antal2006fixation, Blythe07, Cremer11}. The link with the Moran process enables us to take advantage of its well known properties, in particular the exact results for the  fixation probability and mean fixation time~\cite{Moran, Ewens, antal2006fixation, Blythe07}, to characterise analytically many features of our eco-evolutionary model (see sections~\ref{Sec:LargePopulations},~\ref{Sec:Moran}, and annex supplemental section~\ref{SuppSec:DynEnv}).

\subsection{Environmental Fluctuations \& Master Equation}\label{Sec:EnvStat} 
In addition to  demographic fluctuations stemming from random birth and death events, see equation~\eqref{reactions}, we model environmental variability as sudden changes in the available resources, such as in cycles of feast and famine~\cite{Srinivasan98, bernhardt2018metabolic, Merritt18, Himeoka19}. We implement this by letting the carrying capacity be a binary time-fluctuating random variable $K(t)\in\{K_-,K_+\}$, with $K_+>K_-$, as broadly used in eco-evolutionary modelling~\cite{thattai2004stochastic, Bena2006, HL06, hufton2016intrinsic, hidalgo2017species, wienand2017evolution, marrec2020resist, west2020, taitelbaum2020population, Shibasaki2021, taitelbaum2023evolutionary}. This allows us to simply model sudden extreme changes in the population size, particularly the formation of population bottlenecks~\cite{Wahl02, Patwas09, Brockhurst07, Brockhurst07b,Lambert2014,wienand2017evolution}, providing us with a theoretical counterpart of commonly-used laboratory experimental chemostat set-ups~\cite{acar2008, Lambert2014,abdul2021fluctuating, nguyen2021}; see section~\ref{Sec:Discussion2}.

For simplicity, we consider that $K(t)$ is driven by the colored dichotomous Markov noise (DMN) $\xi(t)=\{-1,1\}$ that randomly switches between $K_-$ and $K_+$. The DMN is an important  example of bounded  noise, with finite correlation time, that is easy to simulate accurately (see annex supplemental section~\ref{SuppSec:Num_Sim}) and amenable to analytical progress, and hence often employed in modelling evolutionary processes in fluctuating environments~\cite{Bena2006,HL06,Ridolfi11,hufton2016intrinsic,wienand2017evolution,wienand2018eco,west2020,Shibasaki2021,taitelbaum2020population}. The dynamics of the DMN is defined by the simple reaction~\cite{Bena2006,HL06,Ridolfi11} 

\begin{equation}
 \xi \longrightarrow -\xi, \label{switch}
\end{equation}
endlessly occurring at rate $(1-\delta \xi)\nu$, where $-1<\delta<1$. Here, we always consider the DMN at stationarity where $\xi=\pm 1$ with probability $(1\pm \delta)/2$. The stationary DMN ensemble average is thus $\langle \xi(t)\rangle\equiv \frac{1+ \delta}{2}- \frac{1- \delta}{2}=\delta$ and its auto-covariance (auto-correlation up to a constant) is $\langle \xi(t)\xi(t')\rangle - \langle\xi(t)\rangle\langle\xi(t')\rangle =(1-\delta^2)e^{-2\nu|t-t'|}$,
where $\nu$ is both half the inverse of the correlation time and average switching rate.
We thus consider that the binary switching carrying capacity is~\cite{wienand2017evolution, wienand2018eco, west2020, taitelbaum2020population}
\begin{equation}
K(t)=\frac{1}{2}\left[K_+ +K_- +\xi(t)(K_+ - K_-)\right]
  \label{K(t)},
\end{equation}
and $K(t)$ thus switches from a state where resources are  abundant ($K_+$) to another state where they are scarce ($K_-$) with rates $\nu_+\equiv\nu(1-\delta)$ and $\nu_-\equiv\nu(1+\delta)$ according to \[K_-\xrightleftharpoons[\nu_+]{\nu_-}K_+.\] Environmental statistics can be characterised by the mean switching rate \(\nu\equiv(\nu_-+\nu_+)/2\) and by \(\delta\equiv(\nu_--\nu_+)/(\nu_-+\nu_+)\) that encodes the environmental switching bias: when $\delta>0$, on average, more time is spent in the environmental state $\xi=1$ than  $\xi=-1$, and thus $K=K_+$ is more likely to occur than $K=K_-$ (symmetric switching arises when $\delta=0$). The time-fluctuating carrying capacity \eqref{K(t)} modeling {\it environmental fluctuations} is responsible for the time-variation of the population size, and is coupled with the birth-and-death process \eqref{reactions} and \eqref{eq:transrates}.  

\begin{figure}
\centering
\includegraphics[width=1\textwidth]{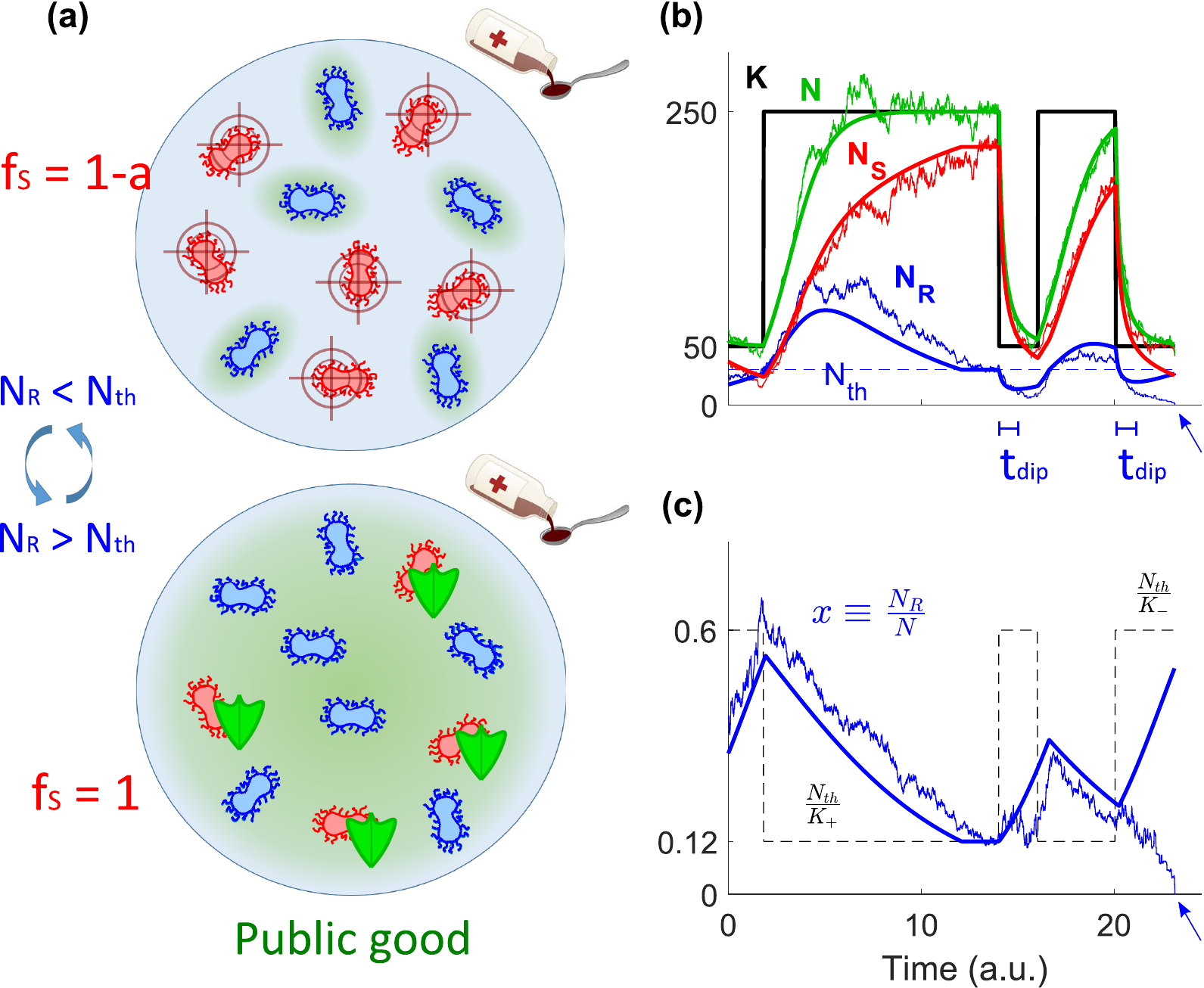}
\caption{\label{fig:SketchAndBehavior}
\textbf{Microbial community model.} {\bf (a)} Top: When the abundance of \(R\) (blue microbes) is below the cooperation threshold \(N_{th}\), antimicrobial drug hinders the growth rate of \(S\) (red microbes) and \(R\) cells have a growth advantage. Bottom: AMR becomes cooperative when the number of \(R\) exceeds \(N_{th}\) and these generate enough {\it resistance enzyme} (public good in green shade) to hydrolyse the antimicrobial drug below the MIC for the whole medium, so that protection against the drug is shared  with \(S\) (with green shields). {\bf (b)} Temporal eco-evolution dynamics of the microbial community for example parameters \(s=0.2\), \(a=0.5\), \(K_{-}=50\), \(K_{+}=250\), \(\nu=0.2\), and \(\delta=0.6\); thick black line shows the sample path of the time-switching carrying capacity \(K(t)\), with a cooperation threshold \(N_{th}=30\) (dashed blue line); thick solid lines depict the \(N\rightarrow\infty\) piecewise deterministic (deterministic between two switches of \(K\)) process defined by equations~\eqref{det_eq_x} and~\eqref{PDMP_N} for the total microbial population (\(N\), green), number of \(R\) (\(N_R=Nx\), blue), and number of \(S\) (\(N_S=N(1-x)\), red); noisy lines show an example stochastic realization of the full model under the joint effect of demographic and environmental fluctuations. In the absence of DN, \(R\) can experience bumps and dips (thick blue line), and $t_{dip}$ indicates the mean time to reach the bottom of a dip from its inception; see section \ref{Sec:TransientDips}. In the presence of DN, fluctuations about the dip can lead to the extinction of \(R\) (blue arrow). {\bf (c)} \(R\) fraction \(x=N_R/N\) for the same sample path of varying environment as in (b); line styles as in  panel (b); the dashed black line shows the stable \(R\) fraction in each environment as \(K(t)\), driven by $\xi(t)$, switches in time.}
\end{figure}

The master equation (ME) giving the probability $P(N_R,N_S,\xi,t)$ for the population to consist of $N_R$ and $N_S$ cells in the environmental state $\xi$ at time $t$ is~\cite{Gardiner}:
\begin{align}
\label{eq:ME}
\hspace{-5mm}
\frac{\partial P(N_R,N_S,\xi,t)}{\partial t} &=  \left( \mathbb{E}_R^--1\right)\left[T^+_R P(N_R,N_S,\xi,t)\right]+\left( \mathbb{E}_S^--1\right)\left[T^+_S P(N_R,N_S,\xi,t)\right] \nonumber \\
&+
\left( \mathbb{E}_R^+-1\right)\left[T^-_R P(N_R,N_S,\xi,t)\right]  +\left( \mathbb{E}_S^+-1\right)\left[T^-_S P(N_R,N_S,\xi,t)\right] \\
&+ \nu_{-\xi} P(N_R,N_S,-\xi,t)-\nu_\xi P(N_R,N_S,\xi,t) \nonumber
\end{align}
where \(\mathbb{E}^{\pm}_{R/S}\) are shift operators such that \(\mathbb{E}^{\pm}_{R/S} f(N_{R/S},N_{S/R},t)=f(N_{R/S}\pm 1,N_{S/R},t)\), and the probabilities are set to \(P(N_R,N_S,\xi,t)=0\) whenever \(N_R<0\) or \(N_S<0\). The last line on the right-hand-side of \eqref{eq:ME} accounts for the random environmental switching, see black line in figure~\ref{fig:SketchAndBehavior}b. Since $T^{\pm}_{R/S}=0$ whenever $N_R=0$ or $N_S=0$, this indicates that there is extinction of $R$ (\(N_R=0\)) and fixation of $S$ (\(N_S=N\)), or fixation of $R$ (\(N_R=N\)) and extinction of $S$ (\(N_S=0\)). When one strain fixates and replaces the other, the population composition no longer changes while its size continues to fluctuate\footnote{The model will finally settle in the absorbing state $N_R=N_S=0$, which corresponds to the eventual extinction of the entire population. This occurs after a time that grows exponentially with the system size and that is unobservable when, as here, $K(t)\gg 1$~\cite{Spalding17, wienand2017evolution, wienand2018eco, taitelbaum2020population}. This phenomenon, irrelevant for our purposes, is not considered here.}. The multivariate ME \eqref{eq:ME} can be simulated exactly using standard stochastic methods~(see annex supplemental section~\ref{SuppSec:Num_Sim}), and encodes the eco-evolutionary dynamics of the model whose main distinctive feature is the \emph{coupling of the population size $N$ and its composition  $x=N_R/N$, with DN coupled to EV}, see \eqref{eq:transrates} and below.

\section{Results}\label{Sec:Results}

In this section we analyse how the coupled demographic and environmental fluctuations shape the evolution of the fraction of $R$ in cooperative AMR~\cite{Coates18}. Our main goals are to establish the conditions under which EV and DN facilitate the eradication of $R$, and reduce the size of the remaining pathogenic microbial population (see also section \ref{Sec:Discussion}).

\subsection{Coupled environmental and demographic noise induces regimes of coexistence and dominance}\label{Sec:PhaseDiagram}

The eco-evolutionary long-lived behavior of a microbial community is chiefly captured by: (I) the expected duration of the strains coexistence (Mean Coexistence Time, MCT,  that here coincides with the unconditional mean fixation  time~\cite{Ewens,antal2006fixation}; see annex section~\ref{SuppSec:MeanCoexTime}); and (II) by the fixation (or extinction) probability of each strain, i.e., the chance that a single strain eventually takes over the entire population (or that the strain is fully replaced by others). These properties have been extensively studied in populations of constant total size, e.g., in terms of the Moran process~\cite{Moran, Gardiner, Ewens, antal2006fixation, Blythe07, traulsen2009stochastic,pinero2022fixation}, but are far less known in communities of fluctuating size when DN is coupled to EV. To gain some insight into the behavior of microbial co-cultures under coupled eco-evolutionary dynamics defined by equation~\eqref{eq:ME}, we compute {\it in silico} the $R$ fixation probability, denoted by \(\phi\), and the strains coexistence probability, labeled by \(P_{\text{coex}}\), when the external conditions fluctuate between harsh (\(K_-=120\), scarce resources) and mild (\(K_+=1000\), abundant resources). Here, \(P_{\text{coex}}\) is defined as the probability that both strains still coexist for a time exceeding twice the average stationary population size \(t>2\langle N\rangle\)\footnote{The rationale is that the MCT in two-strategy evolutionary games scales linearly with $\langle N\rangle$ in neutral regimes, exponentially with $\langle N\rangle$ in coexistence regimes, and sublinearly with $\langle N\rangle$ in regimes where a strain dominates~\cite{antal2006fixation,cremer2009edge,AM10,AM11,AHNRM23}, see also~\cite{reich2007,he2011}. Therefore, \(t>2\langle N\rangle\) is a conservative proxy of coexistence since it allows us to distinguish between regimes where one of the strains dominates and fixates in a time $t\leq 2\langle N\rangle$ from a phase of long-lived coexistence (prior to the eventual fixation of one strain, after a time practically unobservable when $\langle N\rangle \gg 1$).}. In our simulations, we consider a wide range of the switching rate \(\nu\) and bias \(\delta\), with \(\sim10^{3}-10^{4}\) realizations for each dynamic environment, and different values of the cooperation thresholds, with \(N_{th}\sim100\). In our simulations, we respectively use \(s\sim0.1-0.2\) and \(a\sim0.25-0.5\) as plausible values for the resistance metabolic cost and the impact of the drug on $S$~\cite{van2011novo, melnyk2015fitness}. Our choice of $K_{\pm}$ ensures that the dynamics is not dominated mainly by DN or EV, but by the interplay of DN and EV, and the values of the cooperation threshold $N_{th}<K_-$ guarantee that the fixation of either strain or their coexistence are all scenarios arising with finite probabilities in our simulations, see below and annex supplemental section~\ref{SuppSec:Num_Sim}. Note that, as discussed in section~\ref{Sec:Discussion2} and annex section~\ref{SuppSec:BigPop}, the behavior reported here can also be observed in big, realistic populations of \(N>10^{6}\).

\begin{figure}
\centering
\includegraphics[width=1\textwidth]{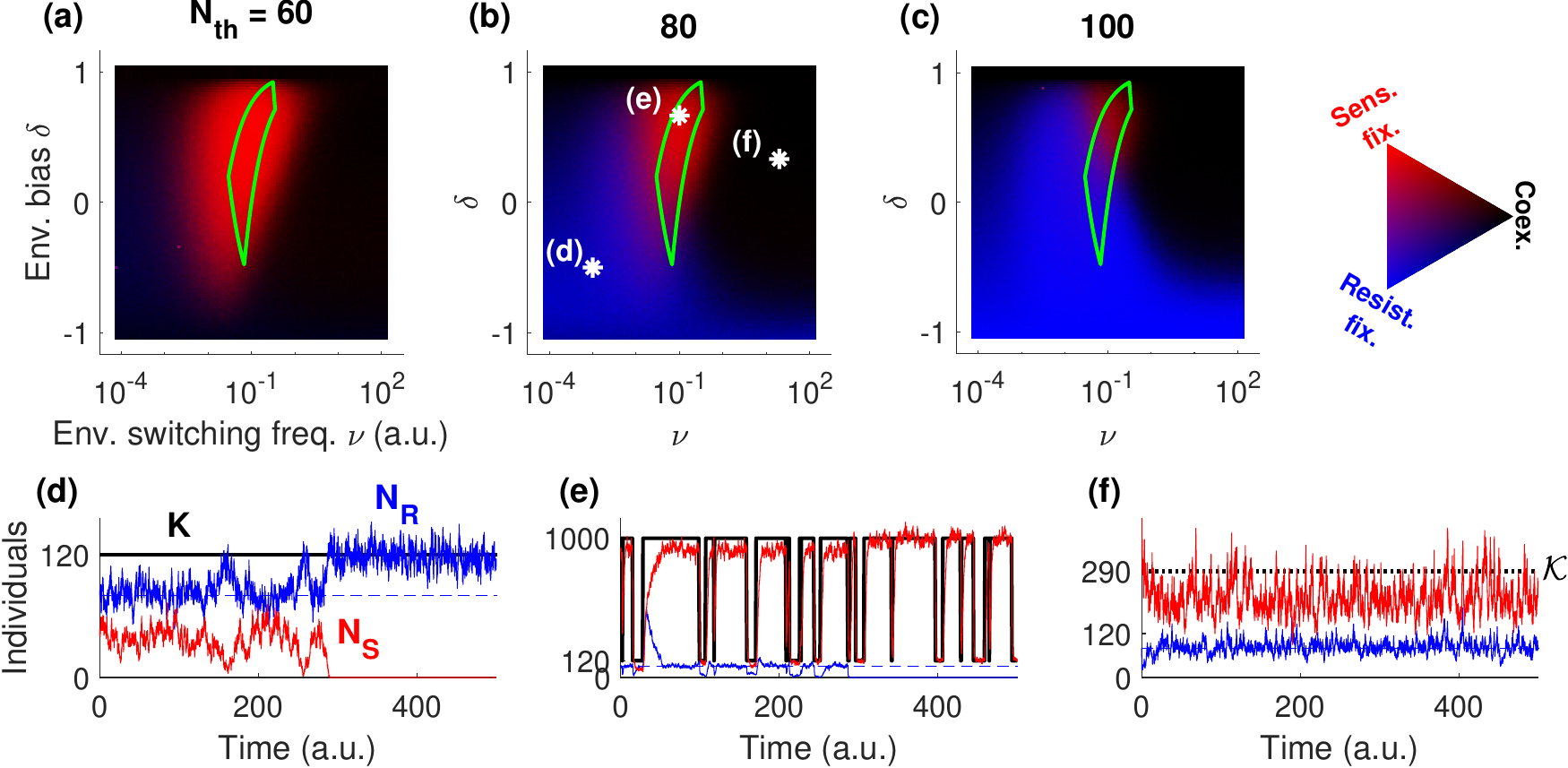}
\caption{\label{fig:PhaseDiagram}\textbf{Eco-evolutionary dynamics in the phase diagram of the joint fixation and coexistence probability.} {\bf (a-c)} Fixation and coexistence joint probability {\it in silico} at a given environmental bias \(\delta\) and mean switching frequency \(\nu\) for \(s=0.1\), \(a=0.25\), \(K_-=120\), and \(K_+=1000\) at resistant cooperation thresholds \(N_{th}=60,~80,\) and \(100\); see the discussions in sections~\ref{Sec:Moran},~\ref{Sec:Discussion2}, and annex supplemental section~\ref{SuppSec:BigPop} for the behavior at much larger populations and thresholds. Stronger blue (red) depicts a higher fixation probability of \(R\) (\(S\)). Darker color indicates higher coexistence probability, defined as the probability to not reach any fixation before \(t=2\langle N\rangle\), where we take the average total population in its stationary state. The area enclosed within the green solid line indicates the optimal regime for the eradication of \(R\), see section~\ref{Sec:TransientDips}. The white asterisks in (b) depict the environmental statistics for each of the bottom panels. {\bf (d-f)} Sample paths for the carrying capacity (\(K\), black), number of \(R\) (\(N_{R}\), blue), number of \(S\) (\(N_{S}\), red), and fixed cooperation threshold \(N_{th}=80\) (dashed blue) for the environmental parameters \(\left(\nu,\delta\right)\) depicted by the corresponding white asterisk in (b). The high environmental switching frequency in (f) results in an effectively constant carrying capacity (\(K=\mathcal{K}\), dotted line); see section~\ref{Sec:LargePopulations}.}
\end{figure}

Figure~\ref{fig:PhaseDiagram}a-c shows the {\it in silico} $\nu \--\delta$ phase diagrams corresponding to the various fixation and coexistence scenarios arising for different cooperation thresholds. For small thresholds relative to EV (\(N_{th}\lesssim10K_{+}/K_{-}\), see annex supplemental section~\ref{SuppSec:BigPop}), $S$ displays a high fixation probability (red region) at intermediate \(\nu\) and non-extreme \(\delta\), where $R$ is most likely to be eradicated. Under high/low values of $\nu$ (when $\delta$ is not too low), the red region in figure~\ref{fig:PhaseDiagram}a-c is surrounded by dark areas where the long-lived coexistence of the strains is most likely. When the threshold \(N_{th}\) is closer to \(K_{-}\), $R$ is most likely to prevail in the blue region of figure~\ref{fig:PhaseDiagram}b-c, where the environment is predominantly in the harsh state ($\delta<0$). As \(N_{th}\) increases, the blue region expands and gradually replaces the red and black areas: the fixation of $R$ is likely to occur in  most of the  $\nu \--\delta$ diagram. In addition to the population makeup, the average population size is a decreasing function of $\nu$ at fixed $\delta$, and increasing with \(\delta\) at fixed \(\nu\); see figure~\ref{fig:TotalPopAndComposition}d-e and section~\ref{Sec:Discussion1}, and \cite{wienand2017evolution,wienand2018eco,taitelbaum2020population,Shibasaki2021,taitelbaum2023evolutionary}.

In what follows, we  analyse the different phases of figure~\ref{fig:PhaseDiagram}a-c, focusing particularly on the characterization of the red area, and also determine how $N_R$ varies with the environmental parameters in the different phases. This allows us to determine \emph{the most favorable environmental conditions  for the eradication of $R$ and for the reduction of the population of pathogenic cells}, which are issues of great biological and practical relevance.

\subsection{Weak demographic noise promotes coexistence}\label{Sec:LargePopulations}
To gain an intuitive understanding of the model's eco-evolutionary dynamics, it is useful to discuss the sample paths of figures~\ref{fig:SketchAndBehavior}b-c and~\ref{fig:PhaseDiagram}d-f in terms of the population size $N$ and the $R$ fraction $x=N_R/N$. 

It is instructive to first consider the case of very large population arising with a constant and large carrying capacity \(K(t)=K_0\gg 1\). In this setting, corresponding to a static environment, we ignore  all forms of fluctuations and  the system evolves according to the mean-field (deterministic) differential equations 
\begin{equation}
    \dot{N}=\sum_{\alpha=R,S}\left(T^+_{\alpha}-T^-_{\alpha}\right)=N\left(1-\frac{N}{K_0}\right), \label{det_eq_N}
\end{equation}
and
\begin{equation}
    \dot{x}=\frac{d}{dt}\frac{N_R}{N}=\frac{T^+_{R}-T^-_{R}}{N}-x\frac{\dot{N}}{N}=\frac{\left(a\theta\left[N_{th}-xN\right]-s\right)x(1-x)}{\left(1-a\theta\left[N_{th}-xN\right]\right)+\left(a\theta\left[N_{th}-xN\right]-s\right)x},
    \label{det_eq_x}
\end{equation}
where the dot indicates the time derivative. It is clear from equation~\eqref{det_eq_x} that the dynamics of the population composition, given by $x$, is coupled to that of its size $N$. According to the logistic equation \eqref{det_eq_N}, the population size reaches $N=K_0$ on a time scale $t\sim 1$ independently of $x$, while the population composition is characterised by a stable equilibrium $x=x_{th}\equiv N_{th}/N=N_{th}/K_0$ reached on a time scale of \(t\sim 1/s\) or \(\sim 1/(a-s)\) from \(x>N_{th}/N\) or \(<N_{th}/N\), respectively. When $s<a\ll1$, there is a timescale separation, with $N$ relaxing to its equilibrium much faster than $x$. We note that the coexistence equilibrium in terms of \(R\) and \(S\) is \(N_{R}^{eq}=N_{th}\) and \(N_{S}^{eq}=K_0-N_{th}\). Clearly, this suggests that $S$ would unavoidably be wiped out if $N_{th}$ was greater than the carrying capacity, and hence we always consider that the latter exceeds the cooperation threshold ($K>N_{th}$).

When the population is large enough for demographic fluctuations to be negligible (\(1/\sqrt{N}\to 0\)) and the sole source of randomness stems from the time-fluctuating environment (random switches of the carrying capacity), the dynamics becomes a so-called piecewise deterministic Markov process (PDMP)~\cite{PDMP}. Between each environmental switch, the dynamics is deterministic and given by equations~\eqref{det_eq_N}, with $K_0$ replaced by $K_{\pm}$ in the environmental state $\xi=\pm 1$, and \eqref{det_eq_x}. Here, the PDMP is thus defined by
\begin{equation}
    \dot{N}=N\left(1-\frac{N}{K(t)}\right)=\begin{cases}
                                            N\left(1-\frac{N}{K_-}\right), & \text{if $\xi=-1$} \\
                                            N\left(1-\frac{N}{K_+}\right), & \text{if $\xi=1$}
                                           \end{cases}
, \label{PDMP_N}
\end{equation}
where the fluctuating carrying capacity \(K(t)\) is given by equation~\eqref{K(t)}, coupled to \eqref{det_eq_x}. Sample paths of this PDMP are shown as solid lines in figures~\ref{fig:SketchAndBehavior}b-c and~\ref{fig:PhaseDiagram}d-f. These realizations illustrate that \(N(t)\) tracks the switching carrying capacity \(K(t)\) independently of \(x\), while \(x(t)\) evolves towards the coexistence equilibrium at the cooperation threshold \(x_{th}(t)=N_{th}/N(t)\), which changes in time as \(N\) varies. Hence, $x$ increases when $N_R<N_{th}$, and it decreases when $N_R>N_{th}$. For extremely high environmental switching rate \(\nu\to\infty\), the microbial community experiences a large number of switches, between any update of the population make-up. In this case, \(N\) is not able to track $K(t)$, but experiences an effectively constant carrying capacity \(K=\mathcal{K}\equiv1/\langle 1/K(t)\rangle\) obtained by self-averaging the environmental noise over its stationary distribution (see \cite{wienand2017evolution,wienand2018eco,west2020,Shibasaki2021,taitelbaum2023evolutionary}), leading to \(\mathcal{K}=2K_+K_-/[(1-\delta)K_++(1+\delta)K_-]\). Hence, when $\nu\to\infty$, the community size is approximately $N\approx \mathcal{K}$ and, provided that $\delta$ is not too close to $-1$ (\(\mathcal{K}\) not too close to \(K_-\)), long-lived coexistence of both strains is likely (with abundances $N_R\approx N_{th}$ and $N_S\approx\mathcal{K}- N_{th}$), as shown in figure~\ref{fig:PhaseDiagram}f.

\subsection{Antimicrobial resistance is robust to demographic noise in static environments}\label{Sec:Moran}

When EV causes a population bottleneck, DN about the coexistence equilibrium may cause the extinction of one strain and the fixation of the other (see figures~\ref{fig:SketchAndBehavior}b-c and~\ref{fig:PhaseDiagram}d-e). To elucidate the fate of microbial communities under fluctuating environments, it is therefore necessary to first understand how a small community is able to fixate, or avoid extinction, in  a static environment, when it is subject to a constant carrying capacity  \(K_0\), with \(1 \ll K_0\sim K_-\ll K_+\). This condition ensures both fixation of one strain or long-lived coexistence are possible, i.e., demographic fluctuations, of order \(\mathcal{O}(1/\sqrt K_0)\), matter but do not govern the dynamics.

Since the community composition tends to the coexistence equilibrium \(x\rightarrow x_{th}\), see equation~\eqref{det_eq_x}, the faster \(N\) dynamics reaches its steady state \(N\to K_0\) before any fixation/extinction events occur, see equation~\eqref{det_eq_N}. Therefore, we assume a fixed \(N=K_0\). The evolutionary dynamics is thus modeled by the analytically tractable Moran process~\cite{Moran,Ewens,Blythe07,Cremer11,wienand2017evolution,wienand2018eco}, where the population composition evolves stochastically by balancing each birth/death of $R$ by the simultaneous death/birth of a $S$, according to the reactions \[N_R+N_S\xrightarrow[]{\widetilde{T}^+_R}(N_R+1)+(N_S-1)\] and \[N_R+N_S\xrightarrow[]{\widetilde{T}^-_R}(N_R-1)+(N_S+1),\] with the effective transition rates $\widetilde{T}^\pm_R=T^\pm_RT^\mp_S/N$ obtained from \eqref{eq:transrates}~\cite{wienand2017evolution,wienand2018eco}.

\begin{figure}
\centering
\includegraphics[width=1\textwidth]{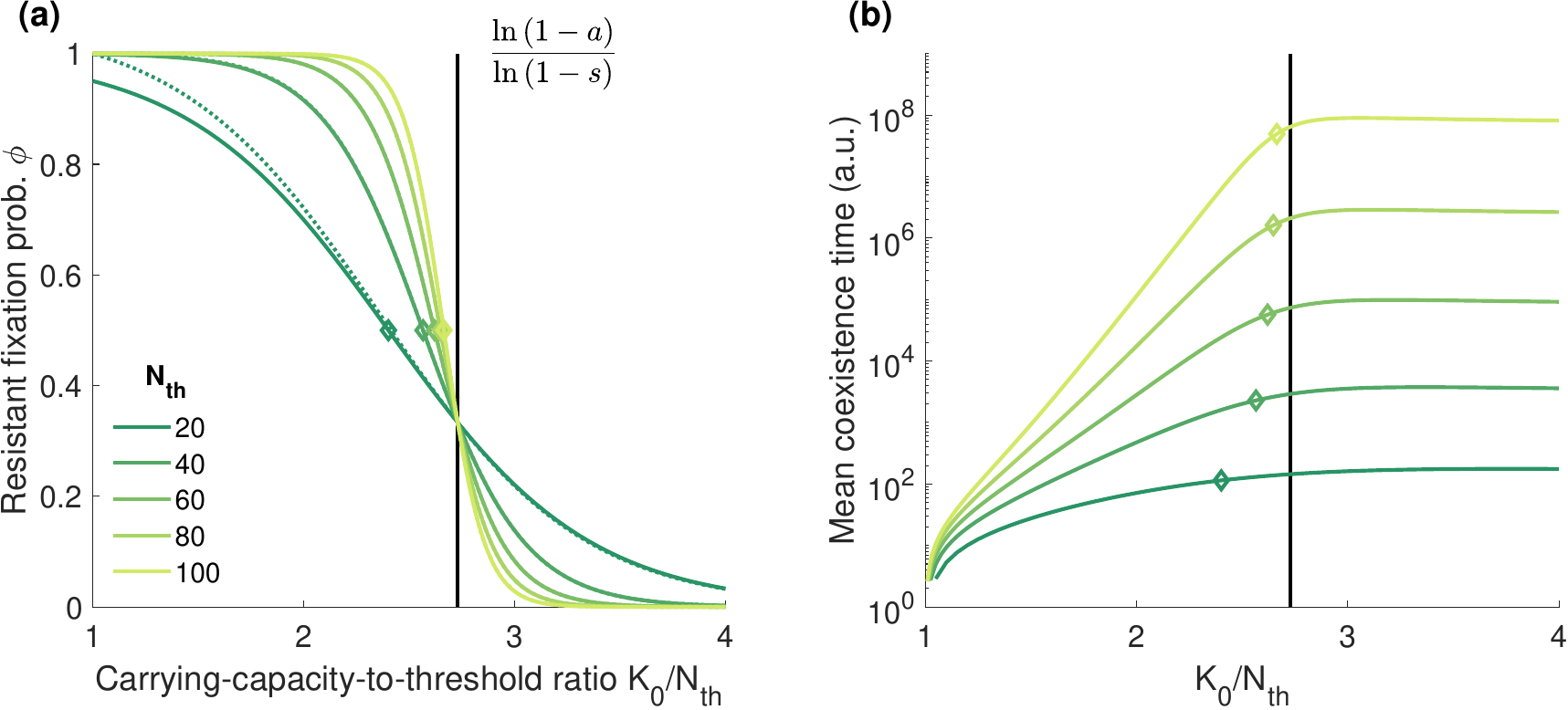}
\caption{\label{fig:FixProbAndMeanAbsTime}
\textbf{Moran theory for \(R\) fixation probability and Mean Coexistence Time (MCT) in static environments.} {\bf (a)} \(R\) fixation probability \(\phi\) in terms of the total microbial population normalised by the resistant cooperation threshold \(K_{0}/N_{th}\) for five example thresholds, from \(N_{th}=20\) (dark green) to 100 (yellow green); the starting microbial composition is set at the coexistence equilibrium \(x_{th}=N_{th}/K_{0}\); solid lines show the approximated prediction of equation~\eqref{eq:ApproxFixProb}; dotted lines depict the exact Moran behavior of annex supplemental equation~\eqref{SuppEq:PartFixProb}, only distinguishable for the smallest threshold; open diamonds illustrate the predicted \(K_{0}^{*}/N_{th}\) that provides fixation equiprobability for each resistant cooperation threshold, see section~\ref{Sec:Moran}. {\bf (b)} Mean Coexistence Time vs \(K_{0}/N_{th}\) in log-linear scale; solid lines show the exact Moran MCT, computed from supplemental equation~\eqref{SuppEq:GenMeanAbsTime}; legend and symbols as in panel (a).}
\end{figure}

Due to DN, the $R$ fraction fluctuates around $x_{th}$ until the eventual extinction of a strain. Therefore, from the classic Moran results (equation~\eqref{SuppEq:PartFixProb} in the annex), we can derive a simplified, approximated expression for the $R$ fixation probability by setting any initial composition directly at coexistence \(x_0=x_{th}\), which yields
\begin{equation}
    \phi\simeq\frac{1}{1+\left(\frac{1}{1-s}\right)^{K_0-K_{0}^{*}}}\text{ ~~with }K_{0}^{*}\equiv N_{th}\frac{\ln\left(1-a\right)}{\ln\left(1-s\right)}-\frac{\ln{\left(\frac{s(1-a)}{(a-s)}\right)}}{\ln{\left(1-s\right)}},
    \label{eq:ApproxFixProb}
\end{equation}
where we now assumed \(\left(1-a\right)^{N_{th}}\ll\left(1-s\right)^{N_{th}}\) and \(\left(1-s\right)^{K_0}\ll\left(1-s\right)^{N_{th}}\), which is in line with our choices $0<s<a<1$ and $N_{th}<K_0$. Here \(K_{0}^{*}\) is the microbial population size giving the same fixation probability \(1/2\) to \(R\) and \(S\). In our examples, \(s=0.1\) and \(a=0.25\) (see section~\ref{Sec:PhaseDiagram}), and fixation equiprobability is reached at \(K_{0}^{*}\approx3N_{th}\), where the \(R\) and \(S\) abundance in the long-lived coexistence equilibrium are respectively \(N_{R}^{eq}\approx K_0/3\) and \(N_{S}^{eq}\approx 2K_0/3\). Figure~\ref{fig:FixProbAndMeanAbsTime}a shows the excellent agreement between the approximation \eqref{eq:ApproxFixProb} (solid lines) and the exact $R$ fixation probability of the underlying Moran process of annex supplemental equation~\eqref{SuppEq:PartFixProb} (dotted lines), for different cooperation thresholds (\(N_{th}=20 \--100\))\footnote{We note that, in the full model simulations at static environments, the total population \(N\) is not fixed but fluctuates about \(K_{0}\). In annex supplemental section~\ref{SuppSec:FluctNFixK} we discuss the minor quantitative impact of these \(N\) fluctuations on the fixation probability and MCT, see supplemental figure~\ref{SuppFig:FixProbAndMeanAbsTime}.}.

Equation~\eqref{eq:ApproxFixProb} and figure~\ref{fig:FixProbAndMeanAbsTime}a-b show that, in a static environment, the relative magnitude of the carrying-capacity-to-threshold ratio \(K_0/N_{th}\) with respect to \(K_{0}^{*}/N_{th}\approx\ln{\left(1-a\right)}/\ln{\left(1-s\right)}\) clearly determines whether \(R\) fixates (for smaller \(K_0/N_{th}\)), becomes extinct (larger \(K_0/N_{th}\)), or coexists with \(S\) for a long time (larger \(K_0/N_{th}\) and large populations). To interpret these results we remember that the mean field behavior tends to \(N_R=N_{th}\) and \(N_S=K_{0}-N_{th}\). When \(K_0/N_{th} \sim 1\), \(N_S\approx K_0-N_{th}\) is small, and $S$ is prone to extinction (\(R\) fixates). As the total population \(K_0\) increases (at fixed cooperation threshold \(N_{th}\)), the equilibrium value \(N_S\approx K_0-N_{th}\) increases, making $S$ less likely to go extinct. The  fixation probability of $S$ thus rises, and overcomes that of the strain \(R\) when \(K_0>K_0^{*}\). However, the expected time for the fixation of \(S\) increases exponentially with \(K_0\) and, for large enough cooperation thresholds (typically for \(N_{th}>50\)), fixation takes too long and is unobservable in practice; see figure~\ref{fig:FixProbAndMeanAbsTime}b. Note that the expected \(S\) fixation time (or \(R\) extinction time) saturates for \(K_0>K_0^{*}\) because the equilibrium value \(N_R\approx N_{th}\) is independent of the total population \(K_0\). For all examples in figure~\ref{fig:PhaseDiagram}a-c we have \(K_0=K_{+}=1000>K_0^*\) when \(\delta=1\) and \(K_0=K_{-}=120<K_0^*\) when \(\delta=-1\). This explains the dark areas (coexistence) in panels b-c where $\delta \to 1$, and the blue regions ($R$ fixation) where $\delta \to -1$. In figure~\ref{fig:PhaseDiagram}a, we observe dark regions (coexistence) for both \(\delta=\pm1\) as the MCT is always larger than the coexistence threshold (\(t>2K_0\)). Therefore, it appears that in static environments AMR always dominates or, at least, survives for extended periods.

\subsection{Demographic noise can eradicate antimicrobial resistance in fluctuating environments}
\label{Sec:TransientDips}
Under low and high environmental switching rates, the community behaves as in static total populations of size $K_{\pm}$ and ${\cal K}$, respectively; see supplemental sections~\ref{SuppSec:SlowEnv}-\ref{SuppSec:FastEnv}. Richer and novel dynamical behavior arises at intermediate switching rate (in figure~\ref{fig:PhaseDiagram}a-c, see red areas around \(\nu=10^{-2}-10^{0}\)), when there are several environmental switches prior to fixation, and the quantities $\phi$ and $P_{\text{coex}}$ cannot be simply expressed in terms of their counterparts in a population of constant effective size. This switching regime is characterised by the full interplay of the ecological and evolutionary dynamics: as shown in figures~\ref{fig:SketchAndBehavior}b and~\ref{fig:PhaseDiagram}e, environmental switches can thus lead to transient ``bumps'' and ``dips'' in $N_R$ (after the carrying capacity increases \(K_{-}\rightarrow K_{+}\) or decreases \(K_{+}\rightarrow K_{-}\), respectively). The transient $N_R$ dips, together with demographic fluctuations caused by the population bottleneck (\(K_+\rightarrow K_-\)), can thus lead to the rapid eradication of $R$ with the fixation of $S$ (red areas in figure~\ref{fig:PhaseDiagram}a-c). Each dip has a small but non-negligible probability to eradicate $R$ and hence reduces the expected coexistence time. Therefore, the ingredients for this fluctuation-driven AMR eradication mechanism are: (I) intermediate environmental switching, so that the total population \(N\) fluctuates by tracking \(K(t)\) without lagging behind, see green lines in figure~\ref{fig:SketchAndBehavior}b; (II) a slower population composition \(x\) coupled to the faster \(N\), so that the \(R\) population \(N_{R}=xN\) experiences transient bumps and dips about its equilibrium \(N_{R}\approx N_{th}\), see blue lines in figure~\ref{fig:SketchAndBehavior}b-c; and (III) a small number of \(R\) at the bottom of transient dips \(N_{R}\sim1\), so that DN can drive \(R\) to extinction, see blue noisy line in figure~\ref{fig:SketchAndBehavior}b.

Here, we are interested in characterising the transient $N_R$ dips as the main fluctuation-driven mechanism leading to the possible eradication of $R$. To study their properties, it is useful to consider the PDMP description of the transient $R$ behavior in large populations:
\begin{equation}
\begin{aligned}
    \dot{N_R}=T^+_{R}-T^-_{R}&=\frac{\left(a-s\right)N_R(\alpha_R-\frac{N_R}{K(t)})}{\left(1-a\right)+\left(a-s\right)N_R/N(t)},&\\
    \text{ with }\alpha_R&\equiv\frac{(1-s)K(t)-(1-a)N(t)}{(a-s)K(t)},&
\end{aligned}
\label{det_eq_Nc}
\end{equation}
where \(K(t)\) and \(N(t)\) are respectively given by equations~\eqref{K(t)} and \eqref{PDMP_N}, and we assume \(N_R<N_{th}\). We note that, after a switch from the mild to harsh environment (\(K_+\rightarrow K_-\)) in the absence of DN, $R$ always survives the ensuing transient dip, and $N_R$ rises towards the coexistence equilibrium \(N_R=N_{th}\); see thick solid line in figure~\ref{fig:SketchAndBehavior}b. However, when \(K_-\ll K_+\) and the microbial community experiences a population bottleneck, a transient dip to a small value of  \(N_R\) can form. When this occurs, $R$ is prone to extinction caused by non-negligible demographic fluctuations (stronger when \(N_R\) is small).
 
To characterise the region of the $\nu \--\delta$ phase diagram where transient $N_R$ dips cause eradication of $R$, we need to estimate $t_{dip}$, defined as the time from the onset of the dip to when $N_R$ reaches its minimal value according to equation~\eqref{det_eq_Nc}, see figure~\ref{fig:SketchAndBehavior}b. To determine $t_{dip}$ from \eqref{det_eq_Nc} we require $\dot{N_R}(t_{dip})=0$ which, assuming  $K_+\gg K_-\gg 1$, yields \(\alpha_R=N_R(t_{dip})/K_{-}\approx0\), implying \(N(t_{dip})\approx K_-(1-s)/(1-a)\). From the solution of equation~\eqref{det_eq_N} with the initial condition \(N(t=0)\approx K_+\), we find:
\begin{equation}
    t_{dip}\approx\ln{\left[\frac{1-s}{a-s}\left(1-\frac{K_-}{K_+}\right)\right]}.
    \label{eq:tdip}
\end{equation}
Ignoring DN, we can thus estimate the \(R\) population at the bottom of the transient dip \(N_{R}^{dip}\), reached at $t=t_{dip}$. This is, we find the \(R\) fraction at the bottom of the dip \(x\left(t_{dip}\right)\) in the small \(x\) limit of equation~\eqref{det_eq_x} and combine it with the above \(N(t_{dip})\) to obtain (see annex supplemental section~\ref{SuppSec:BigPop}):
\begin{equation}
    N_{R}^{dip}=x(t_{dip})N(t_{dip})\approx\frac{N_{th}K_{-}}{K_{+}}\frac{1-s}{1-a}\left(\frac{1-s}{a-s}\right)^{\frac{a-s}{1-a}}\gtrsim\frac{N_{th}K_{-}}{K_{+}},
\label{Eq:betterNdip}
\end{equation}
where we assumed that $R$ started from \(N_R(t=0)=N_{th}\). Demographic fluctuations at the bottom of a dip are of the order $\sqrt{N_{R}^{dip}}$. For DN to possibly drive $R$ to extinction, and the fluctuation-driven eradication scenario to hold, it is necessary that $\sqrt{N_{R}^{dip}}\sim N_{R}^{dip}$, which requires $N_{R}^{dip}=\mathcal{O}(1)$, i.e. $N_{R}^{dip}\sim 10$ or lower. This condition is certainly satisfied when $K_-$ and $N_{th}$  are of comparable size (with $K_->N_{th}$), and each of order \(\sqrt{K_{+}}\), which can also hold for realistically large populations of \(N>10^6\), see section~\ref{Sec:Discussion2} and annex section~\ref{SuppSec:BigPop}.

Under these sufficient requirements, the optimal environmental conditions to rapidly eradicate $R$ in large but fluctuating populations can be estimated from equations~\eqref{det_eq_N}-\eqref{det_eq_x},~\eqref{det_eq_Nc}, and~\eqref{eq:tdip}. First, in the mild  environment (\(K=K_+\)), $R$ needs to be able to evolve to the coexistence equilibrium \(N_R=N_{th}\), requiring a longer average duration of the mild environment \(\nu_{+}^{-1}\) as compared to the evolutionary time scale \(s^{-1}\), i.e., \(\nu_{+}^{-1}\gtrsim s^{-1}\). Second, after the switch from mild to harsh environment (\(K_+\rightarrow K_-\)), $R$ needs to reach the bottom of the transient dip and experience demographic fluctuations, which imposes an average duration of the harsh environment \(\nu_{-}^{-1}\) longer than the average time to reach the bottom of the dip \(t_{dip}\), that is, \(\nu_{-}^{-1}\gtrsim t_{dip}\). Third, if $R$ survives the dip, the environment should go back to the mild $\xi=1$ state to rule out the extinction of $S$ when the environment stays in the harsh state $\xi=-1$. For this, we require the harsh environment to be short, while ensuring that the dip is not interrupted by a switch; see figure~\ref{fig:SketchAndBehavior}b-c. This enforces \(\nu_{-}^{-1}\lesssim2\ln{(\frac{K_{+}}{K_{-}})}(a-s)^{-1}\), where the right-hand-side, derived from the small \(x\) limit of equation~\eqref{det_eq_x}, is twice the expected time to reach the equilibrium in the harsh state \(N_R=N_{th}\) and \(N_S=K_--N_{th}\). As a fourth condition, we demand that this cycle should be as fast as possible to maximise the number of transient dips (while still allowing the population to evolve back to \(N_R=N_{th}\) after a bump), yielding \(\nu_{+}^{-1}\lesssim2\ln{(\frac{K_{+}}{K_{-}})}s^{-1}\), which, similarly as in the previous condition, is twice the average time needed to return to equilibrium in the mild state. Using the environmental parameters \(\nu\) and \(\delta\), the above lead to
\begin{equation}
    \begin{aligned}
        \frac{s}{2\ln{\frac{K_{+}}{K_{-}}}}&\lesssim \nu_{opt}\left(1-\delta_{opt}\right)\lesssim s\text{, and}
        \\
        \frac{a-s}{2\ln{\frac{K_{+}}{K_{-}}}}&\lesssim \nu_{opt}\left(1+\delta_{opt}\right)\lesssim 1/t_{dip}.
    \end{aligned}
    \label{Opt_Env}
\end{equation}
The green contour lines in figure~\ref{fig:PhaseDiagram}a-c enclose the predicted optimal region for the fast eradication of $R$ under \(s=0.1\), \(a=0.25\), \(K_{-}=120\) and \(K_{-}=1000\), and fall in the red areas observed {\it in silico}. The borders of these regions depend on \(N_{th}\). This stems from the dependence of \(\phi\) and MCT on \(N_{th}\) (see figure \ref{fig:FixProbAndMeanAbsTime}b) and the criterion for long-lived coexistence ($t>2\langle N\rangle$). The conservative prediction \eqref{Opt_Env} ignores any dependence on \(N_{th}\). 

In summary, DN can eradicate antimicrobial resistance in fluctuating environments when the population make-up \(x\) evolves on a much slower timescale than the population size \(N\), which requires relatively small values of \(s\) and \(a\). Moreover, the variability in the carrying capacity \(K_+/K_-\) needs to be of the order of the cooperation threshold \(N_{th}\) or larger; the threshold has to fall below the lowest value of the carrying capacity \(K_{-}\); and the switching rate \(\nu\) has to be of order \(s\) and hence comparable to the rate of relaxation of the population composition. Note that all conditions above can be met in biologically relevant systems of any size; see section~\ref{Sec:Discussion2} and annex supplemental section~\ref{SuppSec:BigPop}.

\section{Discussion}\label{Sec:Discussion}
The results of the previous section characterise the long-term microbial population makeup under random switches between mild and harsh environmental conditions (high and low carrying capacity, $K=K_+$ and $K_-$, respectively), for a broad range of the exogenous parameters (mean switching frequency \(\nu\) and switching bias \(\delta\)). Another important aspect of the time evolution of microbial population concerns the nontrivial impact of the environmental variability on the fraction and abundance of drug-resistant ($R$) and drug-sensitive ($S$) microbes in the different regimes, and especially in their phase of coexistence. It is also important to review to what extent our modeling assumptions are amenable to experimental probes. 

\subsection{Impact of environmental variability on the strains fraction and abundance}\label{Sec:Discussion1}

\begin{figure}[h!]
\centering
\includegraphics[width=1\textwidth]{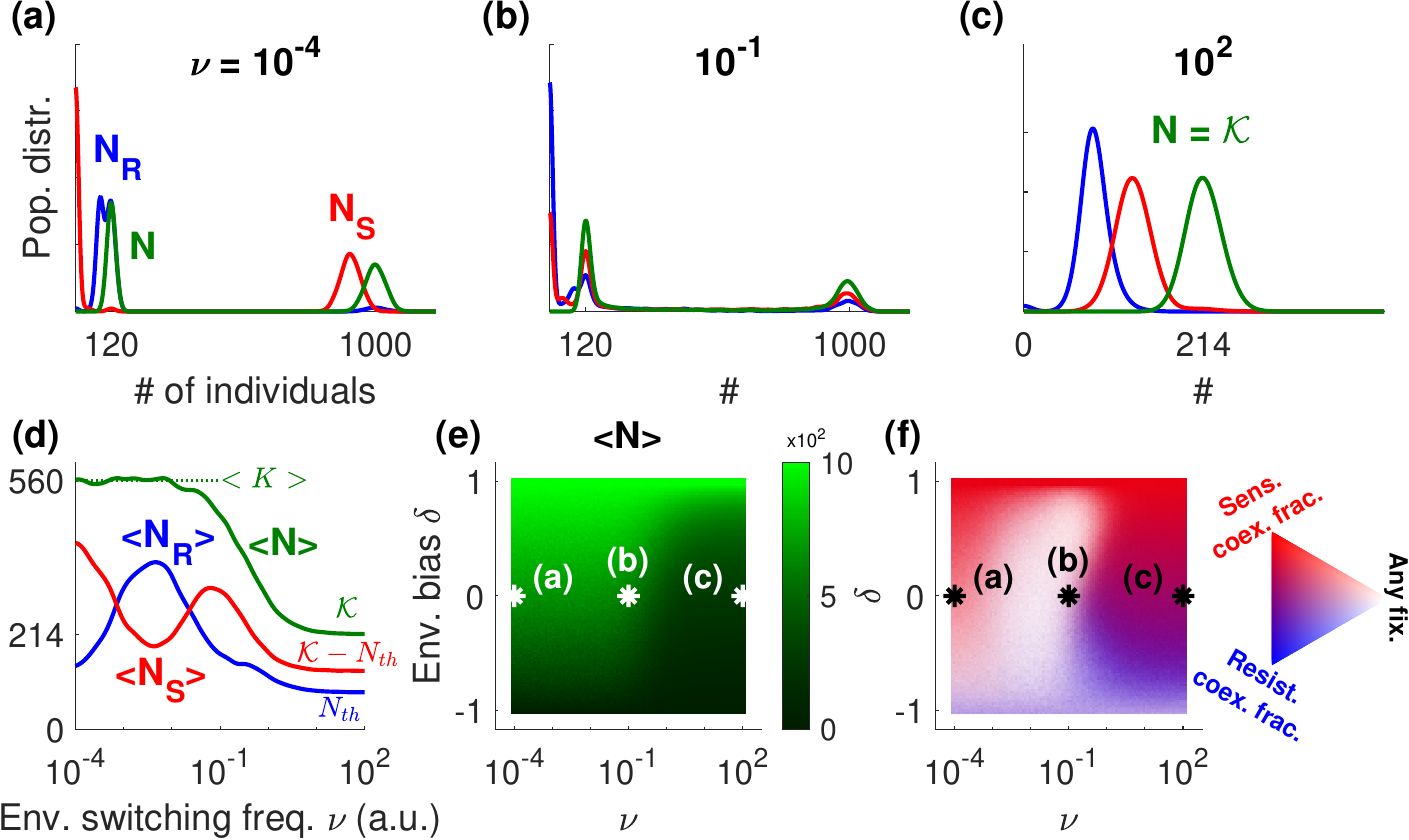}
\caption{\label{fig:TotalPopAndComposition}\textbf{Total population, strain abundance, and coexistence composition in fluctuating environments.} {\bf (a-c)} {\it In silico} probability distributions of the total population ($N$, green), number of $R$ ($N_R$, blue), and number of \(S\) ($N_S$, red), with parameters \(s=0.1\), \(a=0.25\), \(K_-=120\), \(K_+=1000\), and \(N_{th}=80\), under no environmental bias (\(\delta=0\)) and for mean switching rates in slow \(\nu=10^{-4}\) (a), intermediate \(\nu=10^{-1}\) (b), and fast \(\nu=10^{2}\) (c) conditions. Histograms are smoothed by a Gaussian filter of width \(\sigma=10\) in cell number. {\bf (d)} Average overall population (number of individuals on the vertical axis) and strain abundances under no bias, i.e., \(\delta=0\), as a function of switching rate \(\nu\); colors as in (a-c). Lines are smoothed by a log-scale Gaussian filter of width \(\sigma=10\), i.e., one frequency decade. {\bf (e)} Average overall population size in dynamic environments. {\bf (f)} Coexistence composition and fixation probability (of any strain) in dynamic environments. Stronger blue (red) depicts a higher coexistence fraction of \(R\) (\(S\)). Lighter color indicates lower coexistence probability, defined as the probability for no fixation event to occur before \(t=2\langle N\rangle\). The white and black asterisks in (e-f) depict the environmental statistics for each of the top panels. All panels are computed at quasi-stationarity reached after a time \(t=2\langle K\rangle\), ensuring that \(N\) reaches its (quasi-)stationary state, where \(\langle N\rangle\leq\langle K\rangle\); see text.}
\end{figure}
 
It was recently shown that in the fluctuating environment considered here, the average size of the microbial community \(\left<N\right>\) is a decreasing function of the random switching rate $\nu$ (with $\delta$ kept fixed), that \(\left<N\right>\) decreases with lower $\delta$ (keeping $\nu$ fixed), and that $\left<N\right>\to K_\pm$ as $\delta \to \pm1$~\cite{wienand2017evolution, wienand2018eco, taitelbaum2020population, Shibasaki2021}; see figure~\ref{fig:TotalPopAndComposition}d-e. As a consequence, in the blue and red areas of the phase diagrams of  figure~\ref{fig:PhaseDiagram}, where only one strain survives (figure~\ref{fig:PhaseDiagram}a-c), the surviving pathogenic population can be reduced by increasing the environmental switching frequency \(\nu\) and/or the time spent in harsh state (by enforcing \(\delta\rightarrow-1\)). Moreover, since the $R$ fraction \(x\) is directly coupled to \(N\) through the  cooperation threshold \(N_{th}\), see equation~\eqref{det_eq_x}, environmental variability (EV) non-trivially shapes the $R$ fraction in the coexistence regime (colored areas in figure~\ref{fig:TotalPopAndComposition}f).

Under low switching frequency relative to the rate of evolutionary dynamics (see \(\nu\ll\) resistance extra metabolic cost \(s\sim10^{-1}\) in all figures), $R$ cells starting in the mild environment (\(K_{+}\)) are able to reach the coexistence equilibrium \(N_R=N_{th}\) before experiencing a switch; see figure~\ref{fig:SketchAndBehavior}c. However, if the starting environment is harsh (\(K_{-}\)), demographic noise (DN) can rapidly eradicate $S$ and destroy coexistence (see figure~\ref{fig:PhaseDiagram}b-c). The distributions of \(N_R\), \(N_S\) and \(N\) in the regime \(\nu\rightarrow0\) are thus approximately bimodal because they combine both mild and harsh (effectively constant) environments; with \(N_{R}\approx N_{th}\), \(N_{S}\approx K_{+}-N_{th}\), and \(N\approx K_{+}\) for the former; and \(N_{R}\approx K_{-}\), \(N_{S}\approx 0\), and \(N\approx K_{-}\) for the latter; see figure~\ref{fig:TotalPopAndComposition}a. As the switching rate is increased to \(\nu\lesssim s\), fixation dominates, and the \(N_R\) and \(N_S\) bimodal distributions become approximately trimodal, that is, \(N_{R/S}\approx 0\), \(K_{-}\) or \(K_{+}\); see figure~\ref{fig:TotalPopAndComposition}b. The relative weight of the peaks at \(N_{R/S}\approx 0\) is set by \(S/R\) fixation probability, which is modulated by the environmental bias \(\delta\), see supplemental equation~\eqref{SuppEq:SlowFixProb} in the annex. The total population distribution is still bimodal about \(N=K_{\pm}\) since its relaxation dynamics (of timescale \(\sim 1\), see equation~\eqref{det_eq_N}), is faster than the evolutionary timescale $\sim 1/s$. Finally, when \(\nu\) is increased further (\(\nu\gg s\)), we enter the coexistence regime characterised by an effective carrying capacity \(K=\mathcal{K}\)~\cite{wienand2017evolution,wienand2018eco,west2020,taitelbaum2020population,Shibasaki2021,taitelbaum2023evolutionary}, and all distributions become unimodal about the coexistence equilibrium \(N_{R}\approx N_{th}\), \(N_{S}\approx \mathcal{K}-N_{th}\), and \(N\approx \mathcal{K}\); see figure~\ref{fig:TotalPopAndComposition}c.

As a consequence, if $R$ is eradicated, imposing high EV ($\nu\gg s$) and harsh conditions $\delta \to -1$ would considerably reduce the abundance of the surviving community of pathogenic $S$ cells; see figure~\ref{fig:TotalPopAndComposition}d, green solid line, and figure~\ref{fig:TotalPopAndComposition}e. However, if $R$ survives, imposing $\nu\gg 1$ and $\delta<0$ would not only decrease the abundance of both strains but it would also increase the $R$ fraction, and risk further antimicrobial resistance (AMR) spreading, see figure~\ref{fig:TotalPopAndComposition}f (magenta /bluish areas).

\subsection{Review of the modeling assumptions}\label{Sec:Discussion2}
Since we study an idealised microbial model, it is important to 
review our modeling assumptions in light of realistic laboratory experimental conditions. A key assumption to consider is the effectively sharp cooperation threshold \(N_{th}\), which is based on a number of experimental observations of microbial cooperation; see~\cite{davies1994inactivation, Wright05, Yurtsev13, sanchez2013feedback}. Accordingly, we have assumed that EV changes chemical concentrations (e.g., nutrient density) while  the volume of the microbial ecosystem is kept constant~\cite{sanchez2013feedback}. The cooperation threshold is then fixed at a constant number of $R$ microbes \(N_{R}=N_{th}\) because, at constant volume, the resistance enzyme concentration is proportional to the number of public good producers $R$. This crucial ingredient fixes the stable number of $R$ at \(N_{th}\) across fluctuating environments, and is responsible for the transient dips which are at the origin of the novel eco-evolutionary mechanism for the eradication of AMR reported here. The complementary scenario, where the cooperation threshold is set by a fixed $R$ fraction \(x_{th}\) is also relevant (for a different set of microbial ecosystems), and is a topic for future research. Furthermore, in some microbial cases, \(R\) could regulate the production of resistance enzyme by quorum sensing~\cite{pai2012optimality}, but its impact on cooperative AMR remains an open problem. We also note that some resistance mechanisms can show anti-cooperative behaviour, such as efflux-pumps, which could result in an enhanced exposure of sensitive cells to the drug~\cite{poole2007efflux, soto2013role}. In the case of non-shared resistance mechanisms, our model reduces to the eco-evolutionary processes studied in~\cite{wienand2017evolution, wienand2018eco, taitelbaum2020population, Shibasaki2021}. Further analytical results for the non-shared resistance models are discussed in~\cite{uecker2011fixation}, as well as in~\cite{lambert2006probability, parsons2007fixation, patwa2008fixation} in the case of a static environment.

A second assumption to review concerns the simulation results obtained here, for systems with \(K_{\pm}\sim10^{2}-10^{3}\) and \(N_{th}\sim100\), that we are able to computationally probe (see annex supplemental section~\ref{SuppSec:Num_Sim}) but that correspond to populations of relatively small size. In the supplemental section~\ref{SuppSec:BigPop} we provide a detailed discussion on how the rich microbial behavior and novel eco-evolutionary AMR eradication mechanism reported here can be translated to larger, more realistic, microbial communities of size of  order \(N\gtrsim10^{6}\)~\cite{sanchez2013feedback} to  \(N\gtrsim10^{8}\)~\cite{Coates18, smith1956experimental, canetti1956dynamic, feldman1976concentrations, palaci2007cavitary}, or higher. In our discussion we argue that, as long as \(N_{th}K_{-}/K_{+}\lesssim10\) and \(0<s<a\lesssim 10^{-1}-10^{-2}\), regardless of the magnitude of \(K_{\pm}\) or \(N_{th}\), the transient dips studied here will drag $R$ close to extinction, where demographic fluctuations are instrumental for the likely and rapid eradication of AMR. Note that, for very fast/slow fluctuating environments, where transient dips are hindered (see section~\ref{Sec:TransientDips}), \(R\) and \(S\) populations will always coexist unless \(K_{-}-N_{th}\lesssim10\).

A specificity of our study is its focus on biostatic antimicrobial drugs. However, since most antimicrobials gradually change  from acting as biostatic to biocidal as their concentration in the medium grows~\cite{hughes2012selection, andersson2007biological, sanmillan2017fitness}, our approach is consistent with a low antimicrobial concentration scenario. Conveniently, the combined biostatic effect of the drug and the normalization of strain fitness in equation~\eqref{eq:transrates}~\cite{Ewens,Blythe07} decouples the total population \(N\) from its composition \(x\). If any of the above conditions would not hold, \(N\) would then directly depend on \(x\), a case already studied for a simpler model in~\cite{wienand2017evolution, wienand2018eco}. It is worth noting that we have confirmed that the main findings reported here are robust, as they do not depend crucially on the detailed choice of the transition rates in equation~\eqref{eq:transrates}, and in particular they are found to be essentially independent of the normalization by the average fitness; see sections~\ref{Sec:Model} and~\ref{Sec:Moran}. We also note that the values used in our examples for the extra metabolic cost to generate the resistance enzyme (\(s\sim10\%\) to \(25\%\)), and for the impact of the antimicrobial drug on $S$ growth (\(a\sim25\%\) to \(50\%\)), while only indicative, are plausible figures~\cite{van2011novo, melnyk2015fitness}.

For the sake of simplicity, we have focused on modelling EV through binary switches of the carrying capacity. These switches capture sudden changes in the available resources (as in feast and famine cycles~\cite{Srinivasan98, bernhardt2018metabolic, Merritt18, Himeoka19}) that can also occur in presence of antimicrobial drugs, e.g., in polluted environments or during drug treatment. In the context of evolutionary processes,  environments that fluctuate via random switches are commonly  modelled in terms of dichotomous Markov noise (DMN), also known as telegraph noise~\cite{thattai2004stochastic, Bena2006, HL06, hufton2016intrinsic, hidalgo2017species, wienand2017evolution, marrec2020resist, west2020, taitelbaum2020population, Shibasaki2021, taitelbaum2023evolutionary}. Moreover, binary switching is the standard way to implement EV in laboratory-controlled experiments, where the concentration of nutrients can be regulated in a chemostat set-up~\cite{acar2008,Lambert2014,abdul2021fluctuating, nguyen2021}. Although laboratory experiments are often carried out with periodically switching environments (e.g.,~\cite{acar2008,Lambert2014}), and natural environmental conditions often vary continuously in time and magnitude (e.g.,~\cite{nguyen2021}), the relationship between DMN and other commonly used forms of EV has already been extensively studied~\cite{Bena2006, HL06, taitelbaum2023evolutionary}. Therefore, our choice of modelling EV with DMN is natural, convenient and non-limiting: it allows us to make mathematical progress while keeping the theoretical modelling close to laboratory experimental conditions. The literature suggests that the essence of our findings are expected to hold for general fluctuating environments with a time-varying carrying capacity, but the extent to which other and more complex forms of EV than binary random switching may alter our results for microbial communities exhibiting cooperative AMR remains a problem to be studied.

Finally, we note that the novel eco-evolutionary mechanisms reported in this study to eradicate cooperative AMR, and to reduce the total pathogenic microbial community, or minimise the coexistence fraction of $R$, all take place at a biologically and clinically relevant range of environmental switching rates. Indeed, although our theoretical study does not set a specific timescale of microbial growth, a plausible rough estimate for a single replication cycle of a microbe could be of the order of \(\sim1\) hour. The novel AMR eradication mechanism at \(\nu\sim s\) then comes into play when a single environmental phase lasts, on average, \(1/s\sim 10\) hours. This could be consistent with the periodic administration of a treatment that enforces microbial population bottlenecks, and is a feasible time scale for laboratory experiments. Our idealised model however assumes a homeostatic influx of antimicrobial drug in all environments. Thus, an interesting approach for future work would involve the joint application of antimicrobial drug and population bottlenecks (in the harsh environment), with no drug administered in the mild environmental state.

\section{Conclusion}\label{Sec:Conclusions}

Understanding how environmental variability affects the demographic and ecological evolution of microbes is central to tackle the threat of AMR, an issue of pressing societal concern~\cite{oneill2016tackling}. Central questions in studying AMR involve how the fraction of resistant microbes changes in time, and by what mechanisms these can possibly be eradicated. 

It is  well established that AMR is an emergent property of microbial communities, shaped by complex interactions. In particular, certain resistant cells able to inactivate antimicrobials can, under certain conditions, protect the entire microbial community. This mechanism can hence be viewed as an AMR cooperative behavior. Moreover, microbial populations are subject to changing conditions. For instance, the size of a microbial population can vary greatly with the variation of the nutrients or toxins, and can, e.g., experience  bottlenecks. As a result of evolving in volatile environments, microbial communities are prone to be shaped by fluctuations. In general, these stem from environmental variability (EV, exogenous noise) and, chiefly in small populations, from demographic noise (DN). The underlying eco-evolutionary dynamics, characterised by the coupling of DN and EV, is ubiquitous in microbial ecosystems and  plays a key role to understand the AMR evolution, but is still rather poorly understood. 

In this work, we have studied an idealised model of cooperative AMR where a well-mixed, microbial population consisting of sensitive and resistant cells is treated with an antimicrobial (biostatic) drug, hindering microbial growth, in a fluctuating environment. The latter is modeled by a binary switching carrying capacity that fluctuates between two values corresponding to mild and harsh  conditions (high/low values, respectively). Based on a body of experimental work~\cite{Yurtsev13, vega2014collective, meredith2015collective, bottery2016selective,davies1994inactivation, Wright05}, we assume that resistant cells produce, at a metabolic cost, an enzyme that inactivates the antimicrobial drug. Importantly, the abundance of resistant microbes is thus a proxy for the concentration of the drug-inactivating enzyme, which above a certain abundance threshold, becomes a public good by providing drug protection, at no metabolic cost, to the sensitive strain. Above the cooperative threshold, the latter hence have a fitness advantage over the resistant strain, whereas, below the threshold, the drug is responsible for a reduced fitness of the sensitive cells. In this setting, the evolution of AMR can be viewed as a public good problem in a varying environment, whose outcome is shaped by the coupling of environmental and demographic fluctuations.

We have identified three regimes characterising the eco-evolutionary dynamics of the model, associated with the fixation of the resistant or sensitive microbes, or with the long-lived coexistence of both strains. Our analysis shows that, while  AMR generally survives, and often prevails, in static environments, a very different scenario can emerge under environmental variability. In fact, we demonstrate that fluctuations between mild and harsh conditions, coupled to DN, can lead to ``transient dips'' in the abundance of resistant microbes, which can then be driven to extinction by demographic fluctuations. Here, we determine that this \emph{fluctuation-driven AMR eradication mechanism} occurs when the rate of environmental change is comparable to that of the relaxation of the evolutionary dynamics (\(\nu\sim s\)). By computational means, we show that this fluctuation-driven mechanism speeds up the eradication of resistant cells, and argue that it holds also for large microbial communities, comparable to those used in laboratory experiments (\(N>10^6\)). We have also studied how EV non-trivially affects the strain abundance in the various regimes of the model, and in particular have determined  the complex long-lived distribution of the fraction of resistant cells when both strains coexist and the environment fluctuates.

In conclusion, we have shown the existence of a biophysically plausible novel mechanism, driven by the coupling of EV and DN, to eradicate resistant microbes, and have demonstrated how EV shapes the long-lived microbial population in the possible scenarios of strains coexistence or fixation. Our work thus paves the way for numerous possible applications, for instance, in microbial experiments with controlled environmental fluctuations (it is currently possible to track even individual microbes, e.g., see~\cite{bakshi2021tracking, manuse2021bacterial}), which might shed light on new possible treatments against AMR in real-world clinical infections.

\vspace{1cm}

\section*{Data accessibility}
Simulation data and codes for all figures are electronically available from the University of Leeds Data Repository. DOI: https://doi.org/10.5518/1360.

\section*{Author Contributions}
{\bf Llu\'is Hern\'andez-Navarro:} Conceptualization (supporting), Methodology, Formal Analysis (lead), Software, Writing - Original Draft, Writing - Review \& Editing, Visualization, Investigation, Validation. {\bf Matthew Asker:} Formal Analysis (supporting), Software, Writing - Review \& Editing (supporting), Investigation, Validation. {\bf Alastair M. Rucklidge:} Writing - Review \& Editing (supporting), Supervision (supporting), Funding acquisition (supporting). {\bf Mauro Mobilia:} Conceptualization (lead), Methodology (lead), Formal Analysis (supporting), Writing - Original Draft, Writing - Review \& Editing,  Visualization, Supervision (lead), Project administration, Funding acquisition (lead).

Contributor roles taxonomy by CRediT~\cite{brand2015beyond}.

\section*{Competing interests}
We declare we have no competing interests.

\section*{Funding}
L.H.N., A.M.R and M.M. gratefully acknowledge funding from the U.K.
Engineering and Physical Sciences Research Council (EPSRC)
under the grant No. EP/V014439/1
for the project `DMS-EPSRC Eco-Evolutionary Dynamics of Fluctuating Populations’ (https://eedfp.com/). The support of a Ph.D. scholarship to M.A.
by the EPSRC  grant No. EP/T517860/1  is also thankfully acknowledged. 
\section*{Acknowledgements}
We are grateful to K. Distefano, J. Jim\'enez, S. Mu\~{n}oz Montero, M. Pleimling, M. Swailem, and U.C. T\"auber for useful discussions. This work was undertaken on ARC4, part of the High Performance Computing facilities at the University of Leeds, UK.

\bibliographystyle{alpha}
\bibliography{arXiv_main_and_supp}

\newcommand{\etalchar}[1]{$^{#1}$}
\begin{thebibliography}{MHR{\it et al.}21}

\bibitem[AHNRM23]{AHNRM23}
M~Asker, L~Hern\'andez-Navarro, AM~Rucklidge, and M~Mobilia.
\newblock Coexistence of competing microbial strains under twofold
  environmental variability and demographic fluctuations.
\newblock {\em e-print: arXiv}, 2307.06314, 2023.

\bibitem[AM10]{AM10}
M~Assaf and M~Mobilia.
\newblock Large fluctuations and fixation in evolutionary games.
\newblock {\em J. Stat. Mech.}, P09009, 2010.

\bibitem[AM11]{AM11}
M~Assaf and M~Mobilia.
\newblock Large fluctuations and fixation in evolutionary games.
\newblock {\em J. Theor. Biol.}, 275:93, 2011.

\bibitem[AM17]{assaf2017}
M~Assaf and B~Meerson.
\newblock Wkb theory of large deviations in stochastic populations.
\newblock {\em J. Phys. A: Math. Theor.}, 50:263001, 2017.

\bibitem[AMvO08]{acar2008}
M~Acar, J~Mettetal, and A~van Oudenaarden.
\newblock Stochastic switching as a survival strategy in fluctuating
  environment.
\newblock {\em Nat. Genet.}, 40:471--475, 2008.

\bibitem[And07]{anderson2007modified}
DF~Anderson.
\newblock A modified next reaction method for simulating chemical systems with
  time dependent propensities and delays.
\newblock {\em The Journal of Chemical Physics}, 127(21):214107, 2007.

\bibitem[APNK07]{andersson2007biological}
DI~Andersson, SM~Patin, AI~Nilsson, and E~Kugelberg.
\newblock The biological cost of antibiotic resistance.
\newblock {\em Enzyme-Mediated Resistance to Antibiotics: Mechanisms,
  Dissemination, and Prospects for Inhibition}, pages 339--348, 2007.

\bibitem[ARTG21]{abdul2021fluctuating}
F~Abdul-Rahman, D~Tranchina, and D~Gresham.
\newblock Fluctuating environments maintain genetic diversity through neutral
  fitness effects and balancing selection.
\newblock {\em Molecular Biology and Evolution}, 38(10):4362--4375, 2021.

\bibitem[AS06]{antal2006fixation}
T~Antal and I~Scheuring.
\newblock Fixation of strategies for an evolutionary game in finite
  populations.
\newblock {\em Bulletin of Mathematical Biology}, 68(8):1923--1944, 2006.

\bibitem[BAA{\etalchar{+}}15]{brand2015beyond}
A~Brand, L~Allen, M~Altman, M~Hlava, and J~Scott.
\newblock Beyond authorship: attribution, contribution, collaboration, and
  credit.
\newblock {\em Learned Publishing}, 28(2):151--155, 2015.

\bibitem[BBG07]{Brockhurst07}
MA~Brockhurst, A~Buckling, and A~Gardner.
\newblock Population bottlenecks promote cooperation in bacterial biofilms.
\newblock {\em Curr. Biol.}, 17:761, 2007.

\bibitem[Ben06]{Bena2006}
I~Bena.
\newblock Dichotomous markov noise: exact results for out-of-equilibrium
  systems.
\newblock {\em Int. J. Mod. Phys. B}, 20:2825, 2006.

\bibitem[BJ01]{Brown01}
SP~Brown and RA~Johnstone.
\newblock Cooperation in the dark: Signalling and collective action in
  quorum-sensing bacteria.
\newblock {\em Proc. R. Soc. Lond. B. Biol. Sci.}, 268:961, 2001.

\bibitem[BLB{\etalchar{+}}21]{bakshi2021tracking}
S~Bakshi, E~Leoncini, C~Baker, SJ~Ca{\~n}as-Duarte, B~Okumus, and J~Paulsson.
\newblock Tracking bacterial lineages in complex and dynamic environments with
  applications for growth control and persistence.
\newblock {\em Nature Microbiology}, 6(6):783--791, 2021.

\bibitem[BM07]{Blythe07}
RA~Blythe and AJ~McKane.
\newblock Stochastic models of evolution in genetics, ecology and linguistics.
\newblock {\em J. Stat. Mech.}, P07018, 2007.

\bibitem[Bro04]{brook2004beta}
I~Brook.
\newblock $\beta$-lactamase-producing bacteria in mixed infections.
\newblock {\em Clinical Microbiology and Infection}, 10(9):777--784, 2004.

\bibitem[Bro07]{Brockhurst07b}
MA~Brockhurst.
\newblock Population bottlenecks promote cooperation in bacterial biofilms.
\newblock {\em PLoS One}, 2:e634, 2007.

\bibitem[Bro09]{brook2009role}
I~Brook.
\newblock The role of beta-lactamase-producing-bacteria in mixed infections.
\newblock {\em BMC Infectious Diseases}, 9:1--4, 2009.

\bibitem[BSO18]{bernhardt2018metabolic}
JR~Bernhardt, JM~Sunday, and MI~O’Connor.
\newblock Metabolic theory and the temperature-size rule explain the
  temperature dependence of population carrying capacity.
\newblock {\em The American Naturalist}, 192(6):687--697, 2018.

\bibitem[BWB16]{bottery2016selective}
MJ~Bottery, AJ~Wood, and MA~Brockhurst.
\newblock Selective conditions for a multidrug resistance plasmid depend on the
  sociality of antibiotic resistance.
\newblock {\em Antimicrobial Agents and Chemotherapy}, 60(4):2524--2527, 2016.

\bibitem[Can56]{canetti1956dynamic}
G~Canetti.
\newblock Dynamic aspects of the pathology and bacteriology of tuberculous
  lesions.
\newblock {\em American Review of Tuberculosis and Pulmonary Diseases},
  74(2-2):13--21, 1956.

\bibitem[CB06]{Camilli06}
A~Camilli and BL~Bassler.
\newblock Bacterial small-molecule signaling pathways.
\newblock {\em Science}, 311(5764):1113, 2006.

\bibitem[Che94]{Chesson94}
P~Chesson.
\newblock Multispecies competition in variable environments.
\newblock {\em Th.~Pop.~Biol.}, 45:227, 1994.

\bibitem[Che00]{Chesson00}
P~Chesson.
\newblock Mechanisms of maintenance of species diversity.
\newblock {\em Annu. Rev. Ecol. Syst.}, 31:343, 2000.

\bibitem[CK09]{Kimura}
JFF Crow and M~Kimura.
\newblock {\em An Introduction to Population Genetics Theory}.
\newblock Blackburn Press, Caldwell, NJ, USA., 2009.

\bibitem[CMF11]{Cremer11}
J~Cremer, A~Meilbinger, and E~Frey.
\newblock Evolutionary and population dynamics: a coupled approach.
\newblock {\em Phys. Rev. E}, 84:051921, 2011.

\bibitem[CPL{\etalchar{+}}18]{Coates18}
J~Coates, BR~Park, D~Le, \c{S}im\c{s}ek E, W~Chaudhry, and M~Kim.
\newblock Antibiotic-induced population fluctuations and stochastic clearance
  of bacteria.
\newblock {\em eLife}, 7:e32976, 2018.

\bibitem[CRF09]{cremer2009edge}
J~Cremer, T~Reichenbach, and E~Frey.
\newblock The edge of neutral evolution in social dilemmas.
\newblock {\em New Journal of Physics}, 11(9):093029, 2009.

\bibitem[CW81]{Chesson81}
P~L Chesson and R~R Warner.
\newblock Environmental variability promotes coexistence in lottery competitive
  systems.
\newblock {\em The American Naturalist}, 117:923, 1981.

\bibitem[Dav94]{davies1994inactivation}
J~Davies.
\newblock Inactivation of antibiotics and the dissemination of resistance
  genes.
\newblock {\em Science}, 264(5157):375--382, 1994.

\bibitem[Dav84]{PDMP}
MHA Davis.
\newblock Piecewise-deterministic markov processes: a general class of
  non-diffusion stochastic models.
\newblock {\em J. R. Stat. Soc. B}, 46:353, 384.

\bibitem[ESAH19]{Ellner19}
SP~Ellner, RE~Snyder, PB~Adler, and G~Hooker.
\newblock An expanded modern coexistence theory for empirical applications.
\newblock {\em Ecol. Lett.}, 22:3, 2019.

\bibitem[Ewe04]{Ewens}
WJ~Ewens.
\newblock {\em Mathematical Population Genetics}.
\newblock Springer, New York, 2004.

\bibitem[Fel76]{feldman1976concentrations}
WE~Feldman.
\newblock Concentrations of bacteria in cerebrospinal fluid of patients with
  bacterial meningitis.
\newblock {\em The Journal of Pediatrics}, 88(4):549--552, 1976.

\bibitem[Fox13]{Fox13}
JW~Fox.
\newblock The intermediate disturbance hypothesis should be abandoned.
\newblock {\em Trends in Ecology \& Evolution}, 28:86, 2013.

\bibitem[Gar02]{Gardiner}
CW~Gardiner.
\newblock {\em Handbook of Stochastic Methods}.
\newblock Springer, USA, 2002.

\bibitem[GB00]{gibson2000efficient}
MA~Gibson and J~Bruck.
\newblock Efficient exact stochastic simulation of chemical systems with many
  species and many channels.
\newblock {\em The Journal of Physical Chemistry A}, 104(9):1876--1889, 2000.

\bibitem[Gil76]{Gillespie76}
DT~Gillespie.
\newblock A general method for numerically simulating the stochastic time
  evolution of coupled chemical reactions.
\newblock {\em J. Comput. Phys.}, 202:403, 1976.

\bibitem[GPW10]{gaal2010exact}
B~Ga{\'a}l, JW~Pitchford, and AJ~Wood.
\newblock Exact results for the evolution of stochastic switching in variable
  asymmetric environments.
\newblock {\em Genetics}, 184(4):1113--1119, 2010.

\bibitem[Gri73]{Grime73}
JP~Grime.
\newblock Evidence for the existence of three primary strategies in plants and
  its relevance to ecologicaland evolutionary theory.
\newblock {\em Am.~Nat.}, 111:1169, 1973.

\bibitem[HA12]{hughes2012selection}
D~Hughes and DI~Andersson.
\newblock Selection of resistance at lethal and non-lethal antibiotic
  concentrations.
\newblock {\em Current Opinion in Microbiology}, 15(5):555--560, 2012.

\bibitem[HL06]{HL06}
W~Horsthemke and R~Lefever.
\newblock {\em Noise-Induced Transitions}.
\newblock Springer, Berlin, 2006.

\bibitem[HLGM16]{hufton2016intrinsic}
PG~Hufton, YT~Lin, T~Galla, and AJ~McKane.
\newblock Intrinsic noise in systems with switching environments.
\newblock {\em Physical Review E}, 93(5):052119, 2016.

\bibitem[HM20]{Himeoka19}
Y~Himeoka and N~Mitarai.
\newblock Dynamics of bacterial populations under the feast-famine cycles.
\newblock {\em Phys.~Rev.~Research}, 2:013372, 2020.

\bibitem[HMT11]{he2011}
Q~He, M~Mobilia, and UC~T\"auber.
\newblock Coexistence in the two-dimensional {M}ay-{L}eonard model with random
  rates.
\newblock {\em Eur. Phys. J. B}, 82:97–105, 2011.

\bibitem[HSM17]{hidalgo2017species}
J~Hidalgo, S~Suweis, and A~Maritan.
\newblock Species coexistence in a neutral dynamics with environmental noise.
\newblock {\em Journal of theoretical biology}, 413:1--10, 2017.

\bibitem[KA14]{Harrington14}
Harrington KI and Sanchez A.
\newblock Eco-evolutionary dynamics of complex strategies in microbial
  communities.
\newblock {\em Commun. Integr. Biol.}, 7(1):e28230, 2014.

\bibitem[Lam06]{lambert2006probability}
A~Lambert.
\newblock Probability of fixation under weak selection: a branching process
  unifying approach.
\newblock {\em Theoretical population biology}, 69(4):419--441, 2006.

\bibitem[LK14]{Lambert2014}
G~Lambert and E~Kussell.
\newblock Memory and fintess optimization of bacteria under fluctuating
  environments.
\newblock {\em PLoS Genetics}, 10(9):e1004556, 2014.

\bibitem[MB20]{marrec2020resist}
L~Marrec and A-F Bitbol.
\newblock Resist or perish: fate of a microbial population subjected to a
  periodic presence of antimicrobial.
\newblock {\em PLoS Computational Biology}, 16(4):e1007798, 2020.

\bibitem[MHR{\it et al.}21]{Murugan21}
A~Murugan, K~Husain, and MJ~Rust~{\it et al.}
\newblock Roadmap on biology in time varying environments.
\newblock {\em Phys. Biol.}, 18:041502, 2021.

\bibitem[MK18]{Merritt18}
J~Merritt and S~Kuehn.
\newblock Frequency- and amplitude-dependent microbial population dynamics
  during cycles of feast and famine.
\newblock {\em Phys.~Rev.~Lett.}, 121:098101, 2018.

\bibitem[Mor62]{Moran}
PAP Moran.
\newblock {\em The {S}tatistical {P}rocesses of {E}volutionary {T}heory}.
\newblock Oxford, UK: Clarendon, 1962.

\bibitem[MRS11]{Miller11}
AD~Miller, SH~Roxburgh, and K~Shea.
\newblock How frequency and intensity shape diversity-disturbance
  relationships.
\newblock {\em Proc. Natl. Acad. Sci. USA}, 108:5643, 2011.

\bibitem[MSCD{\etalchar{+}}21]{manuse2021bacterial}
S~Manuse, Y~Shan, SJ~Canas-Duarte, S~Bakshi, WS~Sun, H~Mori, J~Paulsson, and
  K~Lewis.
\newblock Bacterial persisters are a stochastically formed subpopulation of
  low-energy cells.
\newblock {\em PLoS Biology}, 19(4):e3001194, 2021.

\bibitem[MSL{\etalchar{+}}15]{meredith2015collective}
HR~Meredith, JK~Srimani, AJ~Lee, AJ~Lopatkin, and L~You.
\newblock Collective antibiotic tolerance: mechanisms, dynamics and
  intervention.
\newblock {\em Nature Chemical Biology}, 11(3):182--188, 2015.

\bibitem[MWK15]{melnyk2015fitness}
AH~Melnyk, A~Wong, and R~Kassen.
\newblock The fitness costs of antibiotic resistance mutations.
\newblock {\em Evolutionary applications}, 8(3):273--283, 2015.

\bibitem[NLGS21]{nguyen2021}
J~Nguyen, J~Lara-Gutiérrez, and R~Stocker.
\newblock Environmental fluctuations and their effects on microbial
  communities, populations and individuals.
\newblock {\em FEMS Microbiol. Rev.}, 45:fuaa068, 2021.

\bibitem[O'N16]{oneill2016tackling}
J~O'Neill.
\newblock Tackling drug-resistant infections globally: final report and
  recommendations.
\newblock {\em Report}, 2016.

\bibitem[PDH{\etalchar{+}}07]{palaci2007cavitary}
M~Palaci, R~Dietze, DJ~Hadad, FKC Ribeiro, RL~Peres, SA~Vinhas, ELN Maciel,
  V~do~Valle~Dettoni, L~Horter, WH~Boom, et~al.
\newblock Cavitary disease and quantitative sputum bacillary load in cases of
  pulmonary tuberculosis.
\newblock {\em Journal of Clinical Microbiology}, 45(12):4064--4066, 2007.

\bibitem[PGH09]{Pelletier09}
F~Pelletier, D~Garant, and HP~Hendry.
\newblock Eco-evolutionary dynamics.
\newblock {\em Phil. Trans. R. Soc. B}, 364:1483, 2009.

\bibitem[Poo07]{poole2007efflux}
K~Poole.
\newblock Efflux pumps as antimicrobial resistance mechanisms.
\newblock {\em Annals of medicine}, 39(3):162--176, 2007.

\bibitem[PQ07]{parsons2007fixation}
TL~Parsons and C~Quince.
\newblock Fixation in haploid populations exhibiting density dependence i: the
  non-neutral case.
\newblock {\em Theoretical population biology}, 72(1):121--135, 2007.

\bibitem[PRS22]{pinero2022fixation}
J~Pi{\~n}ero, S~Redner, and R~Sol{\'e}.
\newblock Fixation and fluctuations in two-species cooperation.
\newblock {\em Journal of Physics: Complexity}, 3(1):015011, 2022.

\bibitem[PTY12]{pai2012optimality}
A~Pai, Y~Tanouchi, and L~You.
\newblock Optimality and robustness in quorum sensing (qs)-mediated regulation
  of a costly public good enzyme.
\newblock {\em Proceedings of the National Academy of Sciences},
  109(48):19810--19815, 2012.

\bibitem[PW08]{patwa2008fixation}
Z~Patwa and LM~Wahl.
\newblock The fixation probability of beneficial mutations.
\newblock {\em Journal of The Royal Society Interface}, 5(28):1279--1289, 2008.

\bibitem[PW09]{Patwas09}
Z~Patwas and LM~Wahl.
\newblock Adaptation rates of lytic viruses depend critically on whether host
  cells survive the bottleneck.
\newblock {\em Evolution}, 64:1166, 2009.

\bibitem[RDL11]{Ridolfi11}
L~Ridolfi, P~D'Odorico, and F~Laio.
\newblock {\em Noise-Induced Phenomena in the Envionmental Sciences}.
\newblock Cambridge University Press, Cambridge, U.K., 2011.

\bibitem[RMF07]{reich2007}
T~Reichenbach, M~Mobilia, and E~Frey.
\newblock Mobility promotes and jeopardizes biodiversity in
  rock–paper–scissors games.
\newblock {\em Nature}, 448:1046–1049, 2007.

\bibitem[Rou79]{Roughgarden79}
J~Roughgarden.
\newblock {\em Theory of Population Genetics and Evolutionary Ecology: an
  Introduction.}
\newblock New York,USA: Macmillan, 1979.

\bibitem[SDF17]{Spalding17}
C~Spalding, CR~Doering, and GR~Flierl.
\newblock Resonant activation of population extinctions.
\newblock {\em Phys. Rev. E}, 96:042411, 2017.

\bibitem[SG13]{sanchez2013feedback}
A~Sanchez and J~Gore.
\newblock Feedback between population and evolutionary dynamics determines the
  fate of social microbial populations.
\newblock {\em PLoS Biology}, 11(4):e1001547, 2013.

\bibitem[SK98]{Srinivasan98}
S~Srinivasan and S~Kjelleberg.
\newblock Cycles of famine and feast: The starvation and outgrowth strategies
  of a marine vibrio.
\newblock {\em J. Biosci.}, 23:501, 1998.

\bibitem[SLKF12]{stegen2012}
JC~Stegen, X~Lin, AE~Konopka, and JK~Fredrickson.
\newblock Stochastic and deterministic assembly processes in subsurface
  microbial communities.
\newblock {\em ISME Journal}, 6:1653--1664, 2012.

\bibitem[SMM17]{sanmillan2017fitness}
A~San~Millan and RC~Maclean.
\newblock Fitness costs of plasmids: a limit to plasmid transmission.
\newblock {\em Microbiology Spectrum}, 5(5):5--5, 2017.

\bibitem[SMM21]{Shibasaki2021}
S~Shibasaki, M~Mobilia, and S~Mitri.
\newblock Exclusion of the fittest predicts microbial community diversity in
  fluctuating environments.
\newblock {\em J. R. Soc. Interface}, 18:20210613, 2021.

\bibitem[Sot13]{soto2013role}
SM~Soto.
\newblock Role of efflux pumps in the antibiotic resistance of bacteria
  embedded in a biofilm.
\newblock {\em Virulence}, 4(3):223--229, 2013.

\bibitem[SPA{\etalchar{+}}12]{shade2012}
A~Shade, H~Peter, S~Allison, D~Baho, M~Berga, H~Buergmann, D~Huber,
  S~Langenheder, J~Lennon, J~Martiny, K~Matulich, T~Schmidt, and Jo~H.
\newblock Fundamentals of microbial community resistance and resilience.
\newblock {\em Frontiers in Microbiology}, 3(417):1--15, 2012.

\bibitem[SWJ56]{smith1956experimental}
MR~Smith and WB~Wood~Jr.
\newblock An experimental analysis of the curative action of penicillin in
  acute bacterial infections: Iii. the efffect of suppuration upon the
  antibacterial action of the drug.
\newblock {\em The Journal of Experimental Medicine}, 103(4):509--522, 1956.

\bibitem[TH09]{traulsen2009stochastic}
A~Traulsen and C~Hauert.
\newblock Stochastic evolutionary game dynamics.
\newblock {\em Reviews of Nonlinear Dynamics and Complexity}, 2:25--61, 2009.

\bibitem[TvO04]{thattai2004stochastic}
M~Thattai and A~van Oudenaarden.
\newblock Stochastic gene expression in fluctuating environments.
\newblock {\em Genetics}, 167(1):523--530, 2004.

\bibitem[TWAM20]{taitelbaum2020population}
A~Taitelbaum, R~West, M~Assaf, and M~Mobilia.
\newblock Population dynamics in a changing environment: random versus periodic
  switching.
\newblock {\em Phys. Rev. Lett.}, 125(4):048105, 2020.

\bibitem[TWMA23]{taitelbaum2023evolutionary}
A~Taitelbaum, R~West, M~Mobilia, and M~Assaf.
\newblock Evolutionary dynamics in a varying environment: Continuous versus
  discrete noise.
\newblock {\em Physical Review Research}, 5(2):L022004, 2023.

\bibitem[UH11]{uecker2011fixation}
H~Uecker and J~Hermisson.
\newblock On the fixation process of a beneficial mutation in a variable
  environment.
\newblock {\em Genetics}, 188(4):915--930, 2011.

\bibitem[VAME10]{visco2010switching}
P~Visco, RJ~Allen, SN~Majumdar, and MR~Evans.
\newblock Switching and growth for microbial populations in catastrophic
  responsive environments.
\newblock {\em Biophysical Journal}, 98(7):1099--1108, 2010.

\bibitem[vdHSS{\etalchar{+}}11]{van2011novo}
MA~van~der Horst, JM~Schuurmans, MC~Smid, BB~Koenders, and BH~ter Kuile.
\newblock De novo acquisition of resistance to three antibiotics by escherichia
  coli.
\newblock {\em Microbial Drug Resistance}, 17(2):141--147, 2011.

\bibitem[VG14]{vega2014collective}
NM~Vega and J~Gore.
\newblock Collective antibiotic resistance: mechanisms and implications.
\newblock {\em Current Opinion in Microbiology}, 21:28--34, 2014.

\bibitem[vK92]{VanKampen}
NG~van Kampen.
\newblock {\em Stochastic Processes in Physics and Chemistry}.
\newblock North-Holland, Amsterdam, 1992.

\bibitem[WFM17]{wienand2017evolution}
K~Wienand, E~Frey, and M~Mobilia.
\newblock Evolution of a fluctuating population in a randomly switching
  environment.
\newblock {\em Phys. Rev. Lett.}, 119(15):158301, 2017.

\bibitem[WFM18]{wienand2018eco}
K~Wienand, E~Frey, and M~Mobilia.
\newblock Eco-evolutionary dynamics of a population with randomly switching
  carrying capacity.
\newblock {\em J. R. Soc. Interface}, 15(145):20180343, 2018.

\bibitem[WGSV02]{Wahl02}
LM~Wahl, PJ~Gerrish, and I~Saika-Voivod.
\newblock Evaluating the impact of population bottlenecks in experimental
  evolution.
\newblock {\em Genetics}, 162:961, 2002.

\bibitem[WM20]{west2020}
R~West and M~Mobilia.
\newblock Fixation properties of rock-paper-scissors games in fluctuating
  populations.
\newblock {\em J. Theor. Biol.}, 491:110135, 2020.

\bibitem[Wri05]{Wright05}
GD~Wright.
\newblock Bacterial resistance to antibiotics: enzymatic degradation and
  modification.
\newblock {\em Adv. Drug Deliv. Rev.}, 57:1451, 2005.

\bibitem[WSWP23]{Plotkin23}
G~Wang, Q~Su, L~Wang, and JB~Plotkin.
\newblock Reproductive variance can drive behavioral dynamics.
\newblock {\em Proc. Natl. Acad. Sci. USA}, 120:e216218120, 2023.

\bibitem[YCD{\etalchar{+}}13]{Yurtsev13}
EA~Yurtsev, HX~Chao, MS~Datta, T~Artemova, and J.~Gore.
\newblock Bacterial cheating drives the population dynamics of cooperative
  antibiotic resistance plasmids.
\newblock {\em Mol. Syst. Biol.}, 9:683, 2013.

\end{thebibliography}

\newpage

\setcounter{figure}{0}
\renewcommand{\figurename}{Figure}
\renewcommand{\thefigure}{S\arabic{figure}}
\setcounter{equation}{0}
\renewcommand{\theequation}{S\arabic{equation}}

\appendix
\title{{\huge\textbf{Appendix: Supplemental Material}}}

\section{Numerical Simulations}\label{SuppSec:Num_Sim}
To study the stochastic behaviour of the microbial community {\it in silico} we have performed {\it exact} stochastic simulations of the underlying birth-death process~\cite{Gillespie76}. Simulations start at an initial time \(t=t_0\) with an initial environment \(K(t_0)\) always at stationarity (with \(\left<\xi(t_0)\right>=\delta\)), initial populations \(N_R(t_0)=N_{th}\) and \(N_S(t_0)=K(t_0)-N_{th}\), and we take into account all the possible reactions that can take place. In the case of the full model this means: (1) the four possible birth or death reactions with rates \{\(T^+_{R}(t_0)\), \(T^-_{R}(t_0)\), \(T^+_{S}(t_0)\), \(T^-_{S}(t_0)\)\} that depend on the variables \{\(N_R(t_0)\), \(N_S(t_0)\), \(K(t_0)\)\} and the constant parameters \{\(s\), \(a\), \(N_{th}\)\}; and (2) the environmental switch with constant rate \(\nu_{\pm}\) for the state \(K(t_0)=K_{\pm}\). We perform efficient stochastic simulations by implementing the Next Reaction Method~\cite{gibson2000efficient} with an improved formulation~\cite{anderson2007modified}. Simulations are run in batches of \(10^3\) realizations for each constant set of parameters \{\(s\), \(a\), \(N_{th}\), \(K_{+}\), \(K_{-}\), \(\nu_{+}\), \(\nu_{-}\), \(K(t_0)\)\}, but for the histograms in main figure~\ref{fig:TotalPopAndComposition}a-c, where we run \(10^{4}\) to get sufficient statistical power.

We choose the lower carrying capacity as \(K_{-}\gg1\) but small enough to capture the impact of DN and bottlenecks on random extinctions of microbial strains; \(K_{-}\ll K_{+}\) different enough so that the environmental changes have a significant impact on the dynamics; and \(K_{+}\) as large as possible to provide insight for more realistic microbial communities, but in a finite computational time. Note that we constrain the analysis to \(N_{th}<K_{-}\) so that microbes tend to a coexistence equilibrium in both environments, see section~\ref{Sec:LargePopulations} in the main text. To estimate the coexistence probability \(P_{\text{coex}}\) we compute the probability that strains fixate only after \(t=2\left<N\right>\) (based on previous works~\cite{cremer2009edge}, see also~\cite{reich2007,he2011}), where the expected total population size is the time average over environmental fluctuations and depends on the statistics \(\nu\) and \(\delta\); see main figure~\ref{fig:TotalPopAndComposition}d-e. We choose this threshold, linear with \(\left<N\right>\), as a conservative proxy to distinguish coexistence and dominance regimes in small populations. The rationale is that the expected duration of coexistence \(t\) in finite two-species populations scales exponentially with the system size \(N\) for the former regime, whereas \(t\) scales logarithmically with \(N\) when there is dominance. The linear case corresponds to the neutral regime~\cite{antal2006fixation,cremer2009edge} and separates the regimes where one species dominate from that where there is a long coexistence of both species.

Main figures~\ref{fig:PhaseDiagram}a-c and~\ref{fig:TotalPopAndComposition}f report diagrams obtained after a time $t=2\langle N\rangle$, where $\langle N\rangle$ is the long-time mean population size. It is useful to notice that the diagrams of these figures have been obtained computationally by letting each simulation run for a time \(\widetilde{t}=2\langle K\rangle=K_++K_- +\delta(K_+-K_-)\), as \(\langle K\rangle\) is also the maximum value that $\langle N \rangle$ can take (for a given fixed $\delta$)~\cite{wienand2017evolution,wienand2018eco,taitelbaum2020population}; see figure~\ref{fig:TotalPopAndComposition}d. We thus record $(N_R(\widetilde{t}), N_S(\widetilde{t}))$, where any \(N_{R/S}(\widetilde{t})=0\) implies fixation of the non-extinct strain, and coexistence corresponds to \(N_{R/S}(\widetilde{t})\neq0\). The histograms in figure~\ref{fig:TotalPopAndComposition}a-c are computed over \(10^{4}\) realizations each, with a Gaussian filter of width \(\sigma=10\) cells to smooth the resulting curves. To computationally obtain the long-time averages of \(N_R\), \(N_S\), and \(N\) of main figure~\ref{fig:TotalPopAndComposition}d-e, we average the triplet $(N_R(\widetilde{t}), N_S(\widetilde{t}), N(\widetilde{t}))$ over \(10^3\) realizations. For panel~\ref{fig:TotalPopAndComposition}d, we apply a Gaussian filter of width \(\sigma=10\), i.e., one decade in the switching frequency log-scale, to smooth the curves. We also note that we obtained the simulation data reported in supplemental figure~\ref{SuppFig:FixProbAndMeanAbsTime} by letting \(10^3\) realizations run until fixation of any strain.

\section{Derivations for the Moran Process}

The Moran process is the stochastic `birth-death' process where the number of individuals of two subpopulations \(R\) and \(S\) evolve at a strictly fixed total number \(N=N_{R}+N_{S}\)~\cite{Moran, Ewens, Blythe07, antal2006fixation, traulsen2009stochastic, wienand2017evolution, wienand2018eco, Cremer11}. The process is fully characterised by the transition rates \(\widetilde{T}^+_R\left(N_R,N\right)\) and \(\widetilde{T}^-_R\left(N_R,N\right)\) that quantify the rate of birth of \(R\) (simultaneously balanced by a single death of \(S\), \(N\) being kept constant) and the rate of death of \(R\) (balanced by a single birth of \(S\)), respectively; see main manuscript section~\ref{Sec:Moran}~\cite{wienand2017evolution, wienand2018eco}. Since the Moran process models populations of constant total size, we set \(N=K_0\), where \(K=K_0\) is the constant carrying capacity in the static environment.

\subsection{Exact General Moran Fixation Probability}\label{A1}

The exact fixation probability $\phi\left(N_R^0,N\right)$ that the subpopulation \(R\) takes over an entire population of size $N$, starting from an initial $R$ number \(N_R^0\), can be derived exactly for a general two-strain Moran model with time-independent transition rates \(\widetilde{T}^{\pm}_R\left(N_R,N\right)\). The exact solution for the fixation probability in the general case is~\cite{Gardiner, VanKampen, Ewens, antal2006fixation, traulsen2009stochastic}
\begin{equation}
    \phi\left(N_R^0,N\right)=\frac{1+\sum_{k=1}^{N_R^0-1}\prod_{i=1}^{k}\gamma\left(i,N\right)}{1+\sum_{k=1}^{N-1}\prod_{i=1}^{k}\gamma\left(i,N\right)},~\text{for}~\gamma\left(N_R^0,N\right)\equiv\frac{\widetilde{T}^-_R\left(N_R^0,N\right)}{\widetilde{T}^+_R\left(N_R^0,N\right)}~\text{and}~N_R^0=1, 2,..., N,
    \label{SuppEq:GenFixProb}
\end{equation}
where the factor \(\gamma\left(N_R,N\right)\) fully determines the above result.

\subsubsection{Exact particular Fixation probability}\label{A1.1}
In our specific model, the effective Moran transition rates in section~\ref{Sec:Moran} of the main manuscript are \(\widetilde{T}^+_R=T^+_RT^-_S/N\) and \(\widetilde{T}^-_R=T^-_RT^+_S/N\), obtained from the main text equation~\eqref{eq:transrates}~\cite{wienand2017evolution,wienand2018eco}, which read
\begin{align}
 \widetilde{T}^+_R\left(N_R,N\right)&=\frac{(1-s)\cdot N_R(N-N_R)/K_0}{1-a\theta\left[N_{th}-N_R\right]+(a\theta\left[N_{th}-N_R\right]-s)N_R/N},\text{~and} \nonumber\\
 \widetilde{T}^-_R\left(N_R,N\right)&=\frac{(1-a\theta\left[N_{th}-N_R\right])\cdot(N-N_R)N_R/K_0}{1-a\theta\left[N_{th}-N_R\right]+(a\theta\left[N_{th}-N_R\right]-s)N_R/N}.
\label{SuppEq:transrates}
\end{align}
In the above transition rates, the constant carrying capacity $K=K_0$ and total population size coincide, in accordance with the tenets of the Moran process, and we therefore set \(N=K=K_0\) in the transition rates \eqref{SuppEq:transrates}. Our particular factor \(\gamma\left(N_R,N\right)\) then yields
\begin{align}
 \gamma\left(N_R\right)&=\frac{1-a\theta\left[N_{th}-N_R\right]}{1-s},
\label{SuppEq:gamma}
\end{align}
which depends piecewise on the number of $R$. Substituting \(\gamma\) in the general exact solution of equation~\eqref{SuppEq:GenFixProb} we get
\begin{equation}
\phi(N_R^0,N)=\begin{cases}
                   0 & N_R^0=0\\
                   
                   \frac{1+\sum_{k=1}^{N_R^0-1}\left(\frac{1-a}{1-s}\right)^k}
                   {1+\sum_{k=1}^{N_{th}-1}\left(\frac{1-a}{1-s}\right)^k+\left(\frac{1-a}{1-s}\right)^{N_{th}-1}\sum_{k=1}^{N-N_{th}}\left(\frac{1}{1-s}\right)^k} & 1\leq N_R^0\leq N_{th}\\
                   \\
                   \frac{1+\sum_{k=1}^{N_{th}-1}\left(\frac{1-a}{1-s}\right)^k+\left(\frac{1-a}{1-s}\right)^{N_{th}-1}\sum_{k=1}^{N_R^0-N_{th}}\left(\frac{1}{1-s}\right)^k}
                   {1+\sum_{k=1}^{N_{th}-1}\left(\frac{1-a}{1-s}\right)^k+\left(\frac{1-a}{1-s}\right)^{N_{th}-1}\sum_{k=1}^{N-N_{th}}\left(\frac{1}{1-s}\right)^k} & N_{th}<N_R^0\leq 1. \\
                \end{cases}
    \label{SpecFixProb2}
\end{equation}
Note that, in consistence with the convention taken in main manuscript's section~\ref{Sec:Model}, we set \(\theta\left[z=0\right]\equiv0\). Making use of the formula for the sum of a finite geometric progression, this becomes
\begin{equation}
\phi(N_R^0,N)=\begin{cases}
                   \frac{1-\left(\frac{1-a}{1-s}\right)^{N_R^0}}
                   {1-\left(\frac{1-a}{1-s}\right)^{N_{th}}+\frac{a-s}{s(1-a)}\left(\frac{1-a}{1-s}\right)^{N_{th}}
                   \left[\left(\frac{1}{1-s}\right)^{N-N_{th}}-1\right]} & 0\leq N_R^0\leq N_{th}\\
                   \\
                   \frac{1-\left(\frac{1-a}{1-s}\right)^{N_{th}}+\frac{a-s}{s(1-a)}\left(\frac{1-a}{1-s}\right)^{N_{th}}
                   \left[\left(\frac{1}{1-s}\right)^{N_R^0-N_{th}}-1\right]}
                   {1-\left(\frac{1-a}{1-s}\right)^{N_{th}}+\frac{a-s}{s(1-a)}\left(\frac{1-a}{1-s}\right)^{N_{th}}
                   \left[\left(\frac{1}{1-s}\right)^{N-N_{th}}-1\right]} & N_{th}<N_R^0\leq 1. \\
                \end{cases}
    \label{SpecFixProb3}
\end{equation}
Finally, the fixation probability of $R$ can be 
written as
\begin{equation}
\phi\left(N_R^0,N,N_{th},s,a\right)=\frac{1-\left(\frac{1-a}{1-s}\right)^{\frac{\left(N_R^0+N_{th}\right)-\left|N_R^0-N_{th}\right|}{2}}+\frac{a-s}{s(1-a)}\left(\frac{1-a}{1-s}\right)^{N_{th}}\left[\left(\frac{1}{1-s}\right)^{\frac{\left(N_R^0-N_{th}\right)+\left|N_R^0-N_{th}\right|}{2}}-1\right]}
    {1-\left(\frac{1-a}{1-s}\right)^{N_{th}}+\frac{a-s}{s(1-a)}\left(\frac{1-a}{1-s}\right)^{N_{th}}\left[\left(\frac{1}{1-s}\right)^{N-N_{th}}-1\right]}.
    \label{SuppEq:PartFixProb}
\end{equation}

An approximate simplification of the above exact result is provided in the main manuscript section~\ref{Sec:Moran} equation~\eqref{eq:ApproxFixProb} by setting \(N_R^0=N_{th}\), \(N=K_{0}\), and assuming \(\left(1-a\right)^{N_{th}}\ll\left(1-s\right)^{N_{th}}\) and \(\left(1-s\right)^{K_0}\ll\left(1-s\right)^{N_{th}}\). Example \(\phi\) values for \(s=0.1\), \(a=0.25\), and several \(N_{th}\) are plotted in main figure~\ref{fig:FixProbAndMeanAbsTime}a and supplemental figure~\ref{SuppFig:FixProbAndMeanAbsTime}a.

\subsection{Exact General Mean Coexistence Time (MCT)}\label{SuppSec:MeanCoexTime}

It is also possible to exactly compute the general mean duration of coexistence regardless of the final state (either fixation or extinction of $R$), i.e., the Mean Coexistence Time (MCT) \(t\left(N_R^0,N\right)\), when the transition rates $\widetilde{T}^\pm_R$ are time-independent. It is worth noting that the MCT here coincides with the unconditional mean fixation time (and mean extinction), since \(t\left(N_R^0,N\right)\) gives the mean time after which coexistence is lost due to the fixation of one strain and the extinction of the other.

The  exact formula for the MCT reads~\cite{Gardiner, VanKampen, Ewens, antal2006fixation, traulsen2009stochastic}:
\begin{equation}
    \begin{aligned}
    t\left(N_R^0,N\right)=-\left[\frac{\sum_{k=1}^{N-1}\sum_{n=1}^{k}\frac{\prod_{m=n+1}^{k}\gamma\left(m,N\right)}{\widetilde{T}^+_R\left(n,N\right)}}{1+\sum_{k=1}^{N-1}\prod_{i=1}^{k}\gamma\left(i,N\right)}\right]\sum_{k=N_R^0}^{N-1}\prod_{i=1}^{k}\gamma\left(i,N\right)+
    \sum_{k=N_R^0}^{N-1}\sum_{n=1}^{k}\frac{\prod_{m=n+1}^{k}\gamma\left(m,N\right)}{\widetilde{T}^+_R\left(n,N\right)},\\
    \text{for}~N_R^0=1, 2,..., N.
    \end{aligned}
    \label{GenMeanAbsTime0}
\end{equation}
where \(\gamma\left(N_R^0,N\right)\) and \(\widetilde{T}^\pm_R\left(N_R^0,N\right)\) are defined as in equations~\eqref{SuppEq:GenFixProb} and~\eqref{SuppEq:transrates}.

For clarity, we split in two the sum over \(k\) in the first numerator (from \(k=1\) to \(N_R^0-1\), and from \(k=N_R^0\) to \(N-1\)), and then rearrange the equation as
\begin{equation}
    \begin{aligned}
    t\left(N_R^0,N\right)=
    \phi\sum_{k=N_R^0}^{N-1}\sum_{n=1}^{k}\frac{\prod_{m=n+1}^{k}\gamma\left(m,N\right)}{\widetilde{T}^+_R\left(n,N\right)}-\left[1-\phi\right]\sum_{k=1}^{N_R^0-1}\sum_{n=1}^{k}\frac{\prod_{m=n+1}^{k}\gamma\left(m,N\right)}{\widetilde{T}^+_R\left(n,N\right)}.
    \end{aligned}
    \label{SuppEq:GenMeanAbsTime}
\end{equation}
where \(\phi\equiv\phi\left(N_R^0,N\right)\) is the $R$ fixation probability starting from \(N_R(t=0)=N_R^0\) in a population of overall size \(N\), and is given by~\eqref{SuppEq:GenFixProb}.

\subsection{Coexistence probability}\label{SuppSec:CoexProb}
To derive an expression for the coexistence probability \(P_{\text{coex}}\) at a fixed total population \(N\), cooperation threshold \(N_{th}\), and starting with the number of $R$ cells at equilibrium \(N_R^0=N_{th}\), we first compute the exact Moran MCT \(t(N_{th},K_{0})\) from equation~\eqref{SuppEq:GenMeanAbsTime}; see main figure~\ref{fig:FixProbAndMeanAbsTime}b and supplemental figure~\ref{SuppFig:FixProbAndMeanAbsTime}b for \(N=K_{0}\), dotted lines. Since the microbial community tends to a coexistence equilibrium, the fixation of a strain occurs on a slow time scale, driven by fluctuations. We thus assume that the full density of coexistence times roughly approximates an exponential distribution of mean \(t(N_{th},K_{0})\), a known property of systems exhibiting metastability~\cite{assaf2017}. \(P_{\text{coex}}\) is thus the exponential cumulative probability remaining after an elapsed time \(2K_{0}\), i.e., \(P\left(t>2\left<N\right>=2K_{0}\right)\), as previously computed {\it in silico} (but for a wide range of \(\nu\)); see main section~\ref{Sec:PhaseDiagram} and figure~\ref{fig:PhaseDiagram}a-c.

\section{Full model in a static environment with constant carrying capacity}\label{SuppSec:FluctNFixK}

In this section we relax the Moran approximation of a fixed total population size, and allow \(N\) to fluctuate around a constant capacity, here denoted by \(K_{0}\). In this case, the behavior of the microbial community does not directly correspond to the Moran process, and follows a bivariate process in terms of the number $N_{R/S}$ of $R/S$ individuals (but it does not depend on $\xi$ since the environment is here static). The probability $P(N_R,N_S,t)$ that the population consists of $N_R$ and $N_S$ at time $t$, now satisfies the ME
\begin{align*}
\label{eq:MEbis}
\hspace{-5mm}
\frac{\partial P(N_R,N_S,t)}{\partial t} &=  \left( \mathbb{E}_R^--1\right)\left[T^+_R P(N_R,N_S,t)\right]+\left( \mathbb{E}_S^--1\right)\left[T^+_S P(N_R,N_S,t)\right] \nonumber \\
&+
\left( \mathbb{E}_R^+-1\right)\left[T^-_R P(N_R,N_S,t)\right]  +\left( \mathbb{E}_S^+-1\right)\left[T^-_S P(N_R,N_S,t)\right].
\end{align*}
where the transition rates $T^{\pm}_{R/S}$ are given by main manuscript, equation~\eqref{eq:transrates}, and the carrying capacity is now constant as $K\to K_0$.

\begin{figure}
\centering
\includegraphics[width=1\textwidth]{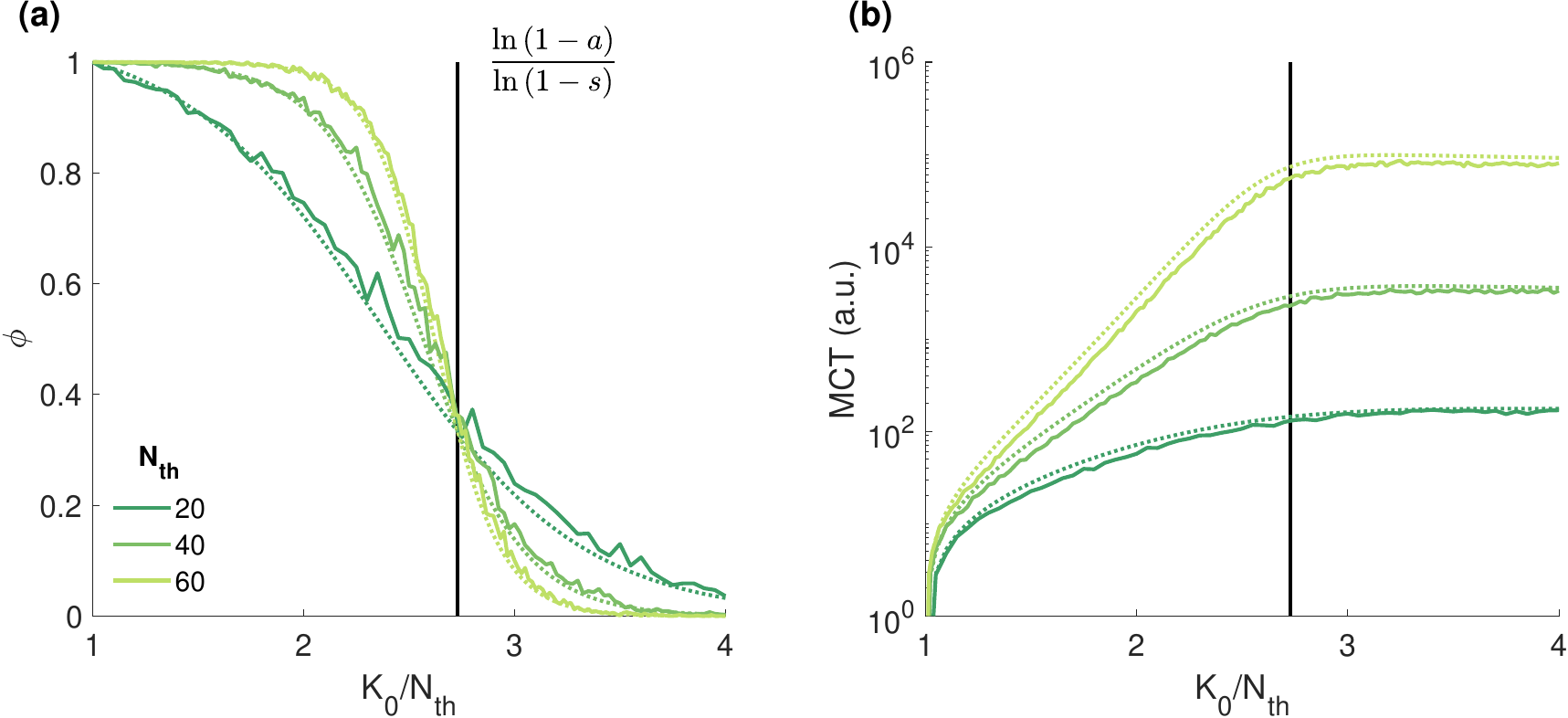}
\caption{\label{SuppFig:FixProbAndMeanAbsTime}
\textbf{Full model simulations in static environments against exact Moran theory for $R$ fixation probability and Mean Coexistence Time (MCT).} {\bf (a)} \(R\) fixation probability \(\phi\) in terms of the total microbial population normalised by the resistant cooperation threshold \(K_{0}/N_{th}\) for three example thresholds, \(N_{th}=20\) (dark green), 40 (green), and 60 (yellow green); the starting microbial composition is set at the coexistence equilibrium \(x_{0}=x_{th}=N_{th}/K_{0}\); dotted lines depict the exact Moran behavior of equation~\eqref{SuppEq:PartFixProb}, solid lines show simulation data of the full model averaged over \(10^3\) realizations. {\bf (b)} Mean Coexistence Time vs \(K_{0}/N_{th}\) in log-linear scale; dotted lines show the exact Moran MCT, computed from equation~\eqref{SuppEq:GenMeanAbsTime}, solid lines show averaged simulation data of the full model over \(10^3\) runs; legend and symbols as in panel (a).}
\end{figure}

For direct comparison, in figure~\ref{SuppFig:FixProbAndMeanAbsTime} we display exact simulations of the above behavior on top of the exact Moran results of main figure~\ref{fig:FixProbAndMeanAbsTime} (dotted lines). We observe that the theoretical predictions of the Moran process quantitatively capture the results {\it in silico} with some minor systematic deviations. These small discrepancies arise from the fact that, below \(K_{0}^*/N_{th}=\ln{\left(1-a\right)}/\ln{\left(1-s\right)}\approx3\) (see main manuscript, section~\ref{Sec:Moran}), the MCT depends exponentially on \(K_{0}/N_{th}\). Therefore, it is exponentially more probable to observe faster fixation under smaller populations, i.e., when random fluctuations drive \(N\) below its expected value \(\left<N\right>=K_{0}\). 

The Moran approximation, based on assuming $N=K_0$, misses demographic fluctuations of order $\sqrt{K_0}$ about $K_0$, which results in underestimating $\phi$ and overestimating the MCT. The relative amplitude of these deviations scale with those of the standard deviations of $N/K_0$ which are of order ${\cal O}(\sqrt{K_0}/K_0)$ and hence become vanishingly small when $K_0\to\infty$.

Hence, the analytical predictions from the Moran process at \(N=K_{0}\) are relevant for static environments because they quantitatively capture both the fixation probability and MCT {\it in silico} within a small relative error. Moreover, this error decreases the bigger the total population and cooperation threshold as the fluctuations about fixed \(K_{0}/N_{th}\) become negligible, which is the case for more realistic, biologically plausible microbial populations.

\section{Dynamic environments: additional analytical derivations}\label{SuppSec:DynEnv}

\subsection{Infrequent environmental switching limit \(\nu\rightarrow0\)}\label{SuppSec:SlowEnv}

When \(\nu\to0\), the average number of environmental switches prior to fixation of one strain (extinction of the other) is very low, and the fixation probability can be obtained by averaging its constant-$N$ counterpart over the stationary distribution of $\xi$. We can indeed  assume that the community evolves subject to a static carrying capacity set by the starting environment \(K(t=0)\equiv K_0\), where \(K_0=K_{\pm}\) with probability \((1\pm\delta)/2\). For these infrequently switching environments, the $R$ fixation probability at an arbitrary environmental bias \(\delta\) is the average
\begin{equation}
\phi\left(\nu\rightarrow0,\delta\right)=\frac{(1+\delta)\phi\left(K_+\right)+(1-\delta)\phi\left(K_-\right)}{2},
\label{SuppEq:SlowFixProb}
\end{equation}
where \(\phi\left(K_{\pm}\right)\equiv\phi\left(N_{th},K_{\pm}\right)\) is the Moran fixation probability of the exact equation~\eqref{SuppEq:PartFixProb}, or approximate main equation~\eqref{eq:ApproxFixProb}, starting at the equilibrium \(N_R^0=N_{th}\). As for the example \(s=0.1\) and \(a=0.25\) parameter values shown in main figures~\ref{fig:PhaseDiagram}-\ref{fig:TotalPopAndComposition}, this \(\phi\left(\nu\rightarrow0,\delta\right)\) $R$ fixation probability is high for \(K_0=K_{-}=120\) with \(\delta=-1\), and gradually (linearly) lowers as \(\delta\) increases and the weighted average shifts towards the starting environment \(K_0=K_{+}=1000\) at \(\delta\to+1\); see figure~\ref{fig:FixProbAndMeanAbsTime}a at the limiting cases \(K_{\pm}/N_{th}\), with \(N_{th}\in[60,100]\); and see the blue-to-black gradient when \(\nu\to0\) for increasing \(\delta\) in figure~\ref{fig:PhaseDiagram}a-c.

Similarly as discussed above for the fixation probability, the coexistence probability at infrequently switching environments averages across the two possible initial environments as
\begin{equation}
    P_{\text{coex}}\left(\nu\rightarrow0,\delta\right) =\frac{(1+\delta)P_{\text{coex}}\left(K_+\right)+(1-\delta)P_{\text{coex}}\left(K_-\right)}{2},
\label{SuppEq:SlowCoexProb}
\end{equation}
where \(P_{\text{coex}}(K_{\pm})\) is the coexistence probability in a static environment \(K_{\pm}\), at fixed total population \(N=K_{\pm}\), and starting $R$ population \(N_R^0=N_{th}\), as derived in the previous supplemental section~\ref{SuppSec:CoexProb}. The slow switching environment coexistence probability \(P_{\text{coex}}\left(\nu\rightarrow0,\delta\right)\) of the above equation~\eqref{SuppEq:SlowCoexProb} is small for low \(\delta\rightarrow -1\) and linearly larger for high \(\delta\rightarrow +1\). This is because the most likely (initial) value of the carrying capacity varies linearly with $\delta$, from  \(K_{0}=K_{-}\) when $\delta\to -1$ to  \(K_{+}\) for $\delta\to 1$; see the MCT in main figure~\ref{fig:FixProbAndMeanAbsTime}b at \(K_{\pm}/N_{th}\) compared to the corresponding coexistence duration threshold \(2K_{\pm}\). Therefore, in dynamic environments with very infrequent switches, microbial behavior shifts from fast fixation (bright) of $R$ (blue) to very slow fixation (black) of $S$, interpreted here as long-lived coexistence (red, overshadowed by black); see section~\ref{Sec:Moran} and figure~\ref{fig:FixProbAndMeanAbsTime} at \(K_{\pm}/N_{th}\). Note that the MCT at \(K\left(\delta\rightarrow-1\right)=K_{-}\) is largest for the smallest threshold \(N_{th}=60\) (see figure~\ref{fig:FixProbAndMeanAbsTime}b), so that the bright blue region in figure~\ref{fig:PhaseDiagram}a is overshadowed by coexistence, in black.

\subsection{Frequent environmental switching limit \(\nu\rightarrow\infty\)}\label{SuppSec:FastEnv}

When \(\nu\rightarrow\infty\), the carrying capacity experiences numerous switches before fixation and extinction occurs. This results in the self-averaging of the environmental noise and the total microbial population tends to the effective carrying capacity ${\cal K}$ seen in main section~\ref{Sec:LargePopulations}~\cite{wienand2017evolution,wienand2018eco,west2020,taitelbaum2020population,Shibasaki2021,taitelbaum2023evolutionary}: \[N\rightarrow\mathcal{K(\delta)}\equiv\frac{2K_+K_-}{(1-\delta)K_++(1+\delta)K_-},\] with $K(\delta\to \pm 1)= K_{\pm}$. Therefore, the theoretical $R$ fixation and coexistence probabilities in the high environmental switching frequency limit are effectively those of a static environment with \(K_{0}=\mathcal{K}\), that is
\begin{equation}
    \phi\left(\nu\rightarrow\infty,\delta\right)=\phi\left(\mathcal{K}(\delta)\right)\text{, and~}P_{\text{coex}}\left(\nu\rightarrow\infty,\delta\right)=P_{\text{coex}}\left(\mathcal{K}(\delta)\right).
\label{SuppEq:FastFixCoexProb}
\end{equation}
where \(\phi\left(\mathcal{K}\left(\delta\right)\right)\) and \(P_{\text{coex}}\left(\mathcal{K}\left(\delta\right)\right)\) are the static environment \(\mathcal{K}\), fixed total population \(N=\mathcal{K}\), starting at equilibrium \(N_R^0=N_{th}\), fixation and coexistence probabilities of the exact equation~\eqref{SuppEq:PartFixProb} (or approximate main equation~\eqref{eq:ApproxFixProb}) and section~\ref{SuppSec:CoexProb}, respectively.

Consistent with the {\it in silico} results of main figure~\ref{fig:PhaseDiagram}a-c with \(\nu\rightarrow\infty\), this limiting behavior also introduces a blue-to-black (nonlinear) gradient for \(\delta=-1\to+1\) as the effective carrying capacity gradually shifts from \(\mathcal{K}=K_{-}\) to \(K_{+}\). As a result, we observe a sharp transition from fixation of $R$ (bright blue) to long-term coexistence (black), with the eventual slow fixation of $S$; see figure~\ref{fig:FixProbAndMeanAbsTime} at \(\mathcal{K}\left(\delta\right)/N_{th}\).

\subsection{Realistic population numbers \(N>10^{6}\)}\label{SuppSec:BigPop}
Typical microbiology laboratory experiments study total microbial populations of typical size \(N\sim10^{6}\) or bigger~\cite{sanchez2013feedback}. These studies  model real-life microbial communities that are usually a few orders of magnitude larger, such as in case studies of mature or chronical clinical infections with \(N\gtrsim10^8\)~\cite{Coates18, smith1956experimental, canetti1956dynamic, feldman1976concentrations, palaci2007cavitary}. In our study, we are constrained to consider systems that are amenable to scrutiny over a wide range of environmental parameters \{\(\nu,\delta\)\} in feasible computational time, and have hence  restricted our {\it in silico} simulations to populations of size up to \(N=1000\). It is thus important to assess analytically how our main findings {\it in silico} translate to  microbial population of more realistic size \(N>10^{6}\).

To this end, we first notice that, as discussed in section~\ref{Sec:Discussion} of the main text, the environmental parameters must fulfill \(1<K_{-}/N_{th}\) to avoid that $S$ dives into extinction following a switch to the harsh environment. For this, the stable number of $S$, \(N_S=K_{-}-N_{th}\) (see section~\ref{Sec:LargePopulations} in the main text), must be high to resist demographic fluctuations, for which a reasonable estimate is \(K_{-}-N_{th}>10\). The second point to notice is that the model tends to two possible coexistence equilibria \(0<x=N_{th}/K_{\pm}<1\) as long as \(K_{-}>N_{th}\) (see section~\ref{Sec:LargePopulations}). Hence, the expected behavior for large populations in the two possible static environments \(K_{\pm}\) is coexistence, which has a duration that scales exponentially with the cooperation threshold \(N_{th}\); see main figure~\ref{fig:FixProbAndMeanAbsTime}b. Since typically the resistant cooperation threshold is of the order \(N_{th}\lesssim K_{-}\), the duration of microbial coexistence is thus expected to grow exponentially with \(K_{-}\). Therefore, fixation in static environments is never observed in realistically big communities where \(K_{-}\gtrsim10^{6}\). 

However, as discussed in the main section~\ref{Sec:TransientDips}, the coupled eco-evolutionary dynamics in fluctuating environments generates significant transient $N_R$ dips when switching from mild to harsh environments (\(K_{+}\rightarrow K_{-}\)) at an intermediate switching rate $\nu\sim s$. The frequency and depth of these transient dips are maximised in a certain range of environmental parameters \(\nu\) and \(\delta\), derived in section~\ref{Sec:TransientDips}. The rapid eradication of $R$ in this optimal dynamic environment regime is shown in the green-enclosing areas of figure~\ref{fig:PhaseDiagram}a-c. We now ask whether realistically big values of \{\(N_{th},K_{-},K_{+}\)\} actually enhance the extinction of cooperative AMR in the fluctuation-driven optimal AMR eradication regime \{\(\nu,\delta\)\}.

In such an optimal AMR-eradication regime, the minimum possible expected number of $R$, reached in the transient dips, is given by the main equation~\eqref{Eq:betterNdip}. To derive this equation, on the one hand we take the low $R$ fraction limit \(x\rightarrow0\) in the main equation~\eqref{det_eq_x} \[\dot{x}\approx\frac{a-s}{1-a}x,\] and thus, assuming \(x(t=0)=x^{eq}_+\equiv\frac{N_{th}}{K_{+}}\), where \(t=0\) is the time at the \(K_{+}\rightarrow K_{-}\) environmental switch, we obtain the microbial composition dynamics at short times \[x(t)\approx\frac{N_{th}}{K_{+}}e^{\frac{a-s}{1-a}t}.\] On the other hand, we can exactly solve the total population logistic dynamics of main equation~\eqref{det_eq_N} in the \(K_{-}\gg1\) environment after the switch, with \(N(0)=K_{+}\), as \[N\left(t\right)=\frac{K_{+}K_{-}e^{t}}{K_{+}\left(e^{t}-1\right)+K_{-}}.\]

Taking \(K_{-}/K_{+}\ll1\) in \(t_{dip}\) given by main equation~\eqref{eq:tdip}; evaluating \(x\left(t=t_{dip}\right)\) above; noting that \(N\left(t=t_{dip}\right)\equiv K_{-}(1-s)/(1-a)\) from $\alpha_R\simeq 0$ in main equation~\eqref{det_eq_Nc}, see main section~\ref{Sec:TransientDips}; and multiplying both resulting expressions, provides the final estimate of the $R$ number \(N_{R}^{dip}\) at the bottom of the transient dip:
\begin{equation}
    N_{R}^{dip}=x\left(t_{dip}\right)N\left(t_{dip}\right)\approx\frac{N_{th}K_{-}}{K_{+}}\frac{1-s}{1-a}\left(\frac{1-s}{a-s}\right)^{\frac{a-s}{1-a}}.
\label{SuppEq:Ndip}
\end{equation}
Since we focus on the case where DN eradicates $R$, for the transient-dip fluctuation-driven eradication mechanism to possibly work, we then need large demographic fluctuations (of order \(\sqrt{N_R^{dip}}\)) relative to \(N_R^{dip}\). This typically suggests to consider \(N_R^{dip}\sim10\) or lower.

Big cooperative AMR microbial communities in ecosystems with antimicrobial drugs could present populations of, for instance, \(N\approx K_{+}\sim10^{12}\) in nutrient abundance conditions. Moreover, biophysically plausible values for the remaining parameters could be \(s=0.1\), \(a=0.25\), and a resistant cooperation threshold \(N_{th}=2\cdot10^{6}\) for an example fixed environmental volume. Sudden and drastic ecological bottlenecks in events of nutrient scarcity (or additional toxins) could then kill most of the community and reduce the total population by a factor of, e.g., few in a million; where the total population would decrease from \(N\sim10^{12}\) to \(N\approx K_{-}\sim5\cdot10^{6}\). These realistic parameters would fulfill the condition \(1<K_{-}/N_{th}=2.5\) with \(K_{-}-N_{th}\sim10^{6}\gg10\). Crucially, these plausible values would fulfill \(N_{R}^{dip}\approx1.7\cdot10\lesssim10\) in equation~\eqref{SuppEq:Ndip}, so that there would be a significant chance that $R$ becomes extinct during each transient $N_R$ dip in dynamic environments.

Finally we note that, as shown above, the relative magnitude of the population bottleneck \(K_{-}/K_{+}\) is critical to enhance the extinction of $R$ during transient dips. Any increase in the population drop between abundance and scarcity environments, i.e., \(K_{-}/K_{+}\rightarrow0\), boosts the eradication of AMR; whereas a reduction in the drop size, i.e., \(K_{-}/K_{+}\rightarrow1\), hinders $R$ extinction and promotes strain coexistence. Tuning the nutrient abundance or scarcity levels in each environment modulates the relative magnitude of the bottleneck \(K_{-}/K_{+}\). But, additionally, introducing an intermediate environmental step between harsh and mild regimes could reduce the population bottleneck and boost coexistence~\cite{sanchez2013feedback}.

\end{document}